\documentclass[prb,superscriptaddress,aps, twocolumn, floats,floatfix, amsmath,amssymb]{revtex4-1}

\usepackage[dvips]{epsfig}
\usepackage{float}
\usepackage{amsmath}
\usepackage{amssymb}
\usepackage{amsfonts}
\usepackage{euscript}
\usepackage{enumerate}
\usepackage{hhline}
\usepackage{threeparttable}
\usepackage{multirow}
\usepackage{supertabular}
\usepackage{pslatex}
\usepackage{tabularx}
\usepackage{color}

\usepackage{graphicx}
\usepackage{dcolumn}
\usepackage{bm}
\usepackage[sort&compress]{natbib} 

\begin{document}

\title{Electron $g$-factor of valley states in realistic silicon quantum dots}

\author{Rusko Ruskov}
\email{ruskovr@lps.umd.edu}
\affiliation{Laboratory for Physical Sciences, 8050 Greenmead Dr., College Park, MD 20740, USA}
\author{Menno Veldhorst}
\affiliation{QuTech and Kavli Institute of Nanoscience, TU Delft, Lorentzweg 1, 2628CJ Delft, The Netherlands}
\author{Andrew S. Dzurak}
\affiliation{Centre for Quantum Computation and Communication
Technology, School of Electrical Engineering and Telecommunications,
The University of New South Wales, Sydney, NSW 2052, Australia}
\author{Charles Tahan}
\email{charlie@tahan.com}
\affiliation{Laboratory for Physical Sciences, 8050 Greenmead Dr., College Park, MD 20740, USA}

\begin{abstract}
We theoretically model the spin-orbit interaction in silicon quantum dot devices,
relevant for quantum computation and spintronics.
Our model is based on a modified effective mass approach
which properly accounts for spin-valley boundary
conditions, derived from the interface symmetry, and should have applicability for other heterostructures.
We show how the valley-dependent interface-induced spin-orbit 2D (3D) interaction,
under the presence of an electric field that is perpendicular to the interface,
leads to a g-factor renormalization in the two lowest valley states of a silicon quantum dot.
These g-factors
can change
with electric field
in opposite direction
when intervalley spin-flip tunneling is favored over intra-valley
processes, explaining recent experimental results.
We show that the quantum dot level structure makes only negligible higher order effects to the g-factor.
We calculate the g-factor as a function of the magnetic field
direction, which is sensitive to the interface symmetry.
We identify spin-qubit dephasing sweet spots at certain directions of the magnetic field, where the
g-factor renormalization is zeroed: these include perpendicular to the
interface magnetic field, and also in-plane directions, the latter being defined
by the interface-induced spin-orbit constants.
The g-factor dependence on
electric field opens the possibility for fast all-electric manipulation of an
encoded, few electron spin-qubit, without the need of a nanomagnet or a nuclear spin-background.
Our approach of an almost fully analytic theory allows for a
deeper physical understanding of the importance of spin-orbit coupling to silicon spin qubits.
\end{abstract}

\maketitle

\section{Introduction}

Electronic $g$-factor arises as a direct consequence of the spin-orbit coupling (SOC);
while relativistic in origin, SOC can be
considerably modified in solids due to
the electron's quasiparticle nature and a non-trivial band structure,
as well as a result of heterostructure confinement effects (see, e.g. Ref.~\onlinecite{RashbaSheka-book1991}).
The variations of  $g$-factor (and more generally, a SOC) in heterostructures and compounds
in externally applied electric or magnetic fields
is at the basis of spintronics
and has led to a multitude of exotic proposals, ranging from spin-transistors \cite{Datta1990}
to topological insulators \cite{Hasan2010}.
While the SOC interaction is often considered in novel materials, it turns out to be a
non-negligible effect in silicon as well \cite{Jansen2012}.
As silicon is recognized as a promising material for spin-based quantum computing \cite{Zwanenburg2013},
understanding the manifiestation and influence of SOC in real devices takes on increased importance.
Particularly relevant are lateral quantum dots (QD) realized in
silicon heterostructures confining few electrons, which allow electric gate control of the spin system
\cite{Maune2012N,Yang2013NC,Xiao2014NC,Kim2014N,Kawakami2014Nn,Veldhorst2014Nn,Veldhorst2015N,
VeldhorstRuskov2015PRB,Delft-experiment,UNSW-experiment-steps,Sandia-experiment1,Grenoble-experiment-ESR-QD1,TanttuDzurak-experiment}.
Silicon can be isotopically enriched to $^{28}$Si and chemically purified,
(see, e.g. Ref.\cite{Itoh2014}), thus removing nuclear spin background as
a major source of spin qubit dephasing.
As a consequence of the increased qubit sensitivity to variations in resonance frequency, 
the $g$-factor's (weak) tunability with an applied electric
field becomes an appreciable tool
for qubit manipulation\cite{Kawakami2014Nn,Veldhorst2014Nn,VeldhorstRuskov2015PRB,Veldhorst2015N}.

\begin{figure}[!h] 
    \centering
        \includegraphics[width=0.62\textwidth]{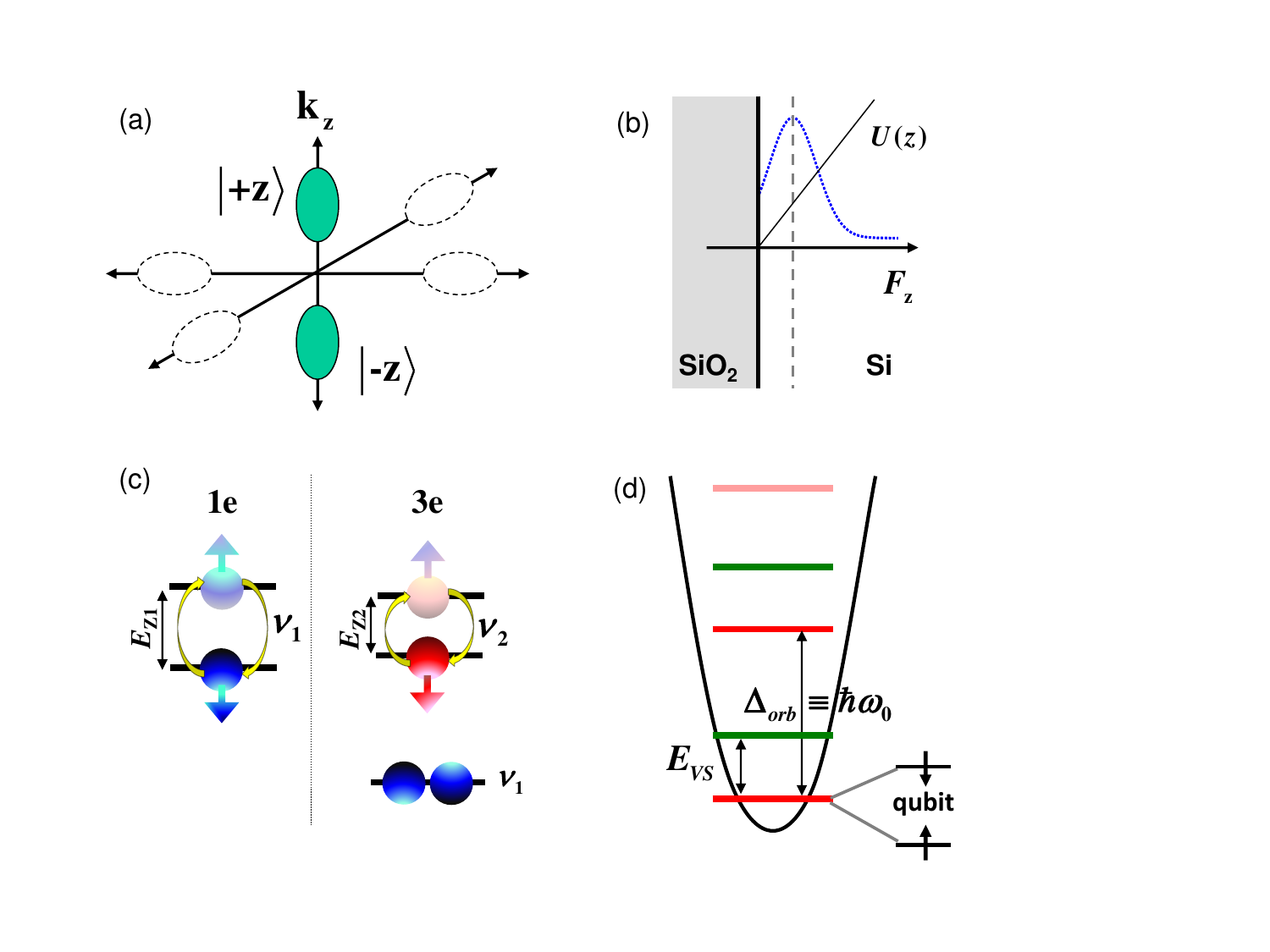}
        \caption{ (a) The six valleys in Silicon. At an  $(0,0,1)$ $\rm Si/SiO_2$ interface (a MOS structure)
        the low-energy subbands are formed by the $\pm \hat{z}$ valleys.
        (b) Confinement in $z$-direction  at the $\rm Si/SiO_2$-interface and with an applied electric
        field $F_z$, forms the eigenvalley states $v_1$, $v_2$, split by a tunable valley splitting $E_{VS} \propto F_z$,
        see Eq.~(\ref{valley_splitting}). Note, that the electron wave function $\varphi(z)$ and its derivative
         $\partial_z \varphi(z)$ may experience a
         discontinuity at the interface region,
         [see Sec.~\ref{Boundary_conditions} and Eq.~(\ref{VolkovPinskerBC1})].
         (c, d): For a small quantum dot the valley splitting is much smaller than the orbital splitting:
         $E_{VS} \ll \Delta_{\rm orb} \equiv \hbar \omega_0$,
         (typically \cite{Veldhorst2014Nn,Yang2013NC,Sandia-experiment1},
         $E_{VS} = 100 - 500\, \mu{\rm eV}$, $\Delta_{\rm orb} = 2 - 8\, {\rm meV}$).
         %
         (c) The one electron g-factor can be approximated by $g_{v_1}$, associated with the lower eigenvalley state $v_1$,
         while the three electron g-factor can be approximated by $g_{v_2}$, associated with the upper eigenvalley $v_2$.
         (d) Higher orbital states only introduce a small second-order effect (Sec. \ref{sec: 2nd-order-g-parallel}), such that
         one is actually measuring just the
         eigenvalley $g$-factors: $g_{1e} \simeq g_{v_1}$ and $g_{3e} \simeq g_{v_2}$.
         }
        \label{fig:1}
\end{figure}

The standard description of the $g$-factor renormalization in a crystal is via a second-order perturbation theory (PT),
using the bulk $\bm{k}\cdot\bm{p}$ Hamiltonian ${\cal H} (\bm{k})$ plus the spin-orbit interaction.
It  is given  as a sum over the virtual electronic excited states (bands),
where a relative contribution of an excited state depends on its coupling to the
electron state of interest via the spin-orbit  interaction Hamiltonian,
and is suppressed  by  the corresponding energy denominator\cite{Roth1960PR}.
In Si, however, the bulk renormalization is very weak (of the order of $\delta g \sim 10^{-3}$),
explained theoretically \cite{Roth1960PR,Liu1962PR} by the  large band-gap   
at the six equivalent conduction-band minima,  at $\bm{k} \approx \hat{n} k_0$,
(with $\hat{n}\equiv \pm \hat{x},\pm\hat{y},\pm\hat{z}$ and $k_0\simeq 0.85\frac{2\pi}{a_0}$), Fig.~\ref{fig:1}a.
A presence of an external electric field $\bm{F}$
only weakly disturbs the crystal symmetry, which leads to even weaker effect for $\delta g(\bm{F})$
(to be discussed below).
In a silicon heterostructure
(in this paper $\rm Si/SiO_2$ is mainly considered as the confinement interface in the growth direction, however the results are
generally applicable to a $\rm Si/Ge$ heterostructures as well),
the band structure is modified due to valley-orbit interaction, reflecting the
reduction of the Si bulk crystal symmetry at the heterostructure interface. This generally leads to
lifting of the six-fold degeneracy:
e.g., for a heterostructure with a growth direction along $[001]$,
four of the valleys are lifted up in energy,
while  at crystal directions $\pm\hat{z}$
a superposition of
the two valley states
forms the lowest eigen-valley states, which are
split-off by the
{\it valley splitting} $E_{\rm VS}$ (Fig.~\ref{fig:1}a and d).   
An applied external electric field, $\bm{F}=(0,0,F_z)$,
enhances the valley splitting, varying in the range of few hundreds $\mu$eV,
which was recently measured in Si quantum dot heterostructures \cite{Yang2013NC,Xiao2014NC}
and confirmed by effective mass
and tight-binding calculations
\cite{OhkawaUemuraI1977,ShamNakayama1979,Chutia2007,Nestoklon3-2008,Saraiva2009PRB}.

It was  stressed by Kiselev {\it et al.}\cite{KiselevIvchenko1992Semicond,KiselevIvchenkoRossler1998PRB}
(see also Refs.\cite{VaskoPrima1981,RodinaEfrosAlekseev2003,Rodina1Alekseev2006})
that  the $g$-factor  renormalization can be equivalently  represented as a first-order
perturbation with the Hamiltonian
$\delta {\cal H} = e \bm{A}\cdot \bm{V}_{\rm bulk}$,
where
$\bm{V}_{\rm bulk} = \hbar^{-1} \partial {\cal H}_{\rm bulk}(\bm{k})/\partial \bm{k}$  is the (bulk) velocity operator,
and  $\bm{A}(\bm{r})$  is the vector potential, which is a linear function of the radius vector $\bm{r}$
for a homogeneous magnetic field.
In low dimensional structures, such as a heterostructure or a quantum well (QW),
this representation is argued to be more effective than the direct PT summation,
leading to the expression for the
$g$-factor tensor ($g_{\alpha\beta}$)\cite{KiselevIvchenkoWillander1997SolStatComm,KiselevIvchenkoRossler1998PRB}:
\begin{equation}
\frac{1}{2} \mu_B \sigma_{\alpha;ss'} g_{\alpha\beta} B_{\beta} \simeq
\frac{1}{2} \mu_B \sigma_{\alpha;ss'} g_0 B_{\alpha}  + \langle e1,s| \delta {\cal H} |e1,s' \rangle
\label{g-factor}  ,
\end{equation}
where $s,s'=\pm 1/2$, $\sigma_{\alpha}$ are the Pauli matrices (for a $1/2$-spinor), and
$|e1,s \rangle$ are the Kramers-conjugate   
lowest subband states.
Given, e.g., an in-plane magnetic field,
the vector potential is $\bm{A} \sim z$, and the matrix element
relates to the ``bulk'' $g$-factor renormalization as:
%
\begin{equation}
\delta g_{\rm bulk} \propto \langle e1,s| \delta {\cal H} |e1,s' \rangle \propto
\langle e1,s| z \bm{V}_{\rm bulk}|e1,s' \rangle  \simeq  \langle z \rangle {\cal V}_{\rm bulk}
\label{bulk-g}  .
\end{equation}
The dependence of $\delta g$ on an external electric field $F_z$
(applied along the growth $z$-direction, as is in the experiment)
may arise from two distinct mechanisms:
(i) from the $z$-confinement deformation of the $\langle z \rangle$ matrix element,
and (ii) from a more subtle mechanism, related to the energy dependence of
the effective mass $m(E)$ 
and other parameters of the bulk $\bm{k}\cdot\bm{p}$   Hamiltonian
(referred  to as non-parabolicity effects: see, e.g. Ref.~\cite{Jancu2005}).

The above, however, is not the whole story. In addition to the
bulk $\bm{k}\cdot\bm{p}$  (effective mass) Hamiltonians
${\cal H}_{\rm bulk}^{A,B}(\bm{k})$  corresponding to the materials $A,B$ that form the heterostructure,
there is also an interface region (with size of the order of the materials' lattice constants, $a_A,a_B$).
The latter can be described to a good approximation with an
energy-independent transfer matrix $\hat{T}_{\rm if}$ that characterizes
solely the interface region
(see, e.g.
Refs.~\onlinecite{Rodina1Alekseev2006,VolkovPinsker1979,Vasko-book1998,AndoMori82,Tokatly2002,Braginsky1999}),
and relates the wave functions and
their derivatives, $\Psi_{A,B}^n$, $\partial_z \Psi_{A,B}^n$,
at the interface
(see Fig.~\ref{fig:1}b and the discussion below);
here, $n$ enumerates the bands (and their degeneracies) in each material.
The transfer matrix $\hat{T}_{\rm if}$ amounts to a certain boundary
condition on the (envelope) wave function
components $\Psi_{A,B}^n$, $\partial_z \Psi_{A,B}^n$, which can be equivalently expressed
as an interface Hamiltonian ${\cal H}_{\rm if}(\bm{k})$.
Thus, one arrives at an ``interface'' $g$-factor renormalization of the form:
\begin{equation}
\delta g_{\rm if} \propto \langle z \rangle {\cal V}_{\rm if}
\label{interface-g}  ,
\end{equation}
where ${\cal V}_{\rm if}$ is a ``velocity'' associated with the
interface Hamiltonian\cite{VolkovPinsker1979,VaskoPrima1981,Vasko-book1998,DevizorovaVolkov2014}.
We argue in what follows that in a $\rm Si/SiO_2$-inversion layer the
interface mechanism dominates the
bulk, $\delta g_{\rm if} \gg \delta g_{\rm bulk}$.
Physically, the interface contribution is expected to be large for quite
distinctive materials
such as $\rm Si/SiO_2$;
however, it cannot be excluded {\it a priori} in less distinctive heterostructures,
e.g., in GaAs/AlGaAs or Si/Ge ones.

This paper is a thorough study of the theoretical construction
and its consequences that was
suggested in our original short paper publication \cite{VeldhorstRuskov2015PRB}.
Results include
{\it general models of the valley splitting,
valley-dependent SOC interactions,
and valley-dependent anisotropic g-factors at a Si-heterostructure interface.}
In particular,
(1)  We obtain an interface modified effective mass approach where the electron spin and valley components
are mixed at the heterostructure interface via a non-trivial boundary condition (BC),
in the presence of a perpendicular electric field, Sec. II. 
This BC is equivalent to intervalley tunneling  plus
intervalley and intra-valley electron spin-flip processes,
and reflects the interface $C_{\rm 2v}$ symmetry.
The derived interface Hamiltonian is singular (in the heterostructure growth $z$-direction),
which does not allow simple perturbation theory (PT) for the $g$-factor. 

(2)  We obtain from the BC a smooth interface 3D SOC tunneling Hamiltonian
(Sec. III A)
that allows PT for the $g$-factor renormalizations  while maintaining the gauge invariance of the results.
From the interface Hamiltonian we derive  the electric field dependent
valley splitting at the Si  heterostructure, Sec. III B, for a general interface-confinement potential,
allowing us to interpret the experiment of Ref.~\onlinecite{Yang2013NC}.

(3)  In the spin-valley mixing sector we obtain, in a translationally invariant form,
the  valley-diagonal Rashba and Dresselhaus effective 2D  SOC Hamiltonians,
as well as the off-diagonal in eigenvalleys Rashba and Dresselhaus SOCs, Sec. III C.
The corresponding valley-dependent Rashba and Dresselhaus SOC constants  
for a linear $z$-confinement scale linearly with the electric field, $\propto F_z$,
as does the valley splitting.
%
The valley dependencies of the SOC constants suggest they may change sign when one switches
between eigenvalleys,
as a {\it consequence of  the dominance of the intervalley spin-flipping processes vs. the intravalley process}.

(4) The valley-dependent $g$-factor tensor renormalizations for an in-plane magnetic field
are derived
in Sec. IV B from the smooth interface 3D SOC Hamiltonians,
scaling as $\propto F_z^{2/3}$ for a linear $z$-confinement.
For a perpendicular magnetic field, the relevant $g$-factor tensor components scale linearly
with $F_z$, Sec. IV C,  being proportional to the non-vanishing electric dipole matrix elements
(scf. Refs.~\cite{Yang2013NC,Gamble2013PRB}).

(5) We show that the  sign change of the SOC constants for different eigenvalleys
leads to a corresponding sign change
of the $g$-factor renormalization.
In particular, for the in-plane magnetic field in a $[110]$-direction, we derive qualitatively
and quantitatively that the  $g$-factor renormalization is opposite in sign for an electron
occupying different eigenvalley states, Fig.~\ref{fig:1}c, as it was observed in the experiment
\cite{VeldhorstRuskov2015PRB}, Sec. IV B.

(6)  A prediction is made for the  $g$-factor  angular dependence on the in-plane magnetic field,
as well as for an out-of-plane magnetic field
in Sec.~IV B, C, and D,
that is in accordance with the $C_{\rm 2v}$ interface symmetry,
which was confirmed in current experiments \cite{Sandia-experiment1,TanttuDzurak-experiment}.
The $g$-factor angular dependence
provides a
single QD spin qubit with decoherence sweet spots with respect to
the magnetic field direction.

(7)
In Secs. IV B and C we consider
second order corrections to the $g$-factor
originating
from the QD internal level structure, Fig.~\ref{fig:1}d, also including the effect of interface roughness \cite{Yang2013NC}.
For both the in-plane and perpendicular magnetic field configurations,
these corrections (for a Si QD with strong lateral confinement)
can be neglected: $\delta^{(2)} g \sim 10^{-6}$.

(8)
Finally, in Sec. IV E, we compare our results to various current experiments \cite{VeldhorstRuskov2015PRB,Sandia-experiment1},
providing in particular estimations for the ratio   
of the lower eigenvalley SOC constants,
as well as for the difference of the SOC constants in both eigenvalleys subspaces
with the account for the $g$-factor offsets for each eigenvalley.
%
The dephasing mechanism introduced by the $g$-factor electric field dependence,
is in a qualitative agreement with the experiment \cite{VeldhorstRuskov2015PRB}.
The results of Sec. IV  can be seen as an experimental proposal to better understand
the spin-valley structure at a Si interface.
Section V contains the summary of results,
and a discussion related to recent experiments with MOS QD structures \cite{Sandia-experiment1}.
More details of the derivations are presented in Appendices A, B, C.

\section{$\rm Si/SiO_2$ interface and boundary conditions}

\subsection{Valley and spin scattering at a $\rm Si/SiO_2$ heterostructure}

We will consider a $\rm Si/SiO_2$ heterostructure grown  along the $[001]$ ($\hat{z}$) direction
with Si at $z>0$ under an applied electric field in the $\hat{z}$-direction,
$(0,0,F_z)$ corresponding to a linear  
 potential $U_z(z)=|e| F_z z$.
Due to a large conduction band offset to $\rm SiO_2$ ($\Delta_{\rm offset} \approx 3\,{\rm eV}$)
we will approximate it with an infinite boundary, $U_z(z)=\infty,\, z<0$ (Fig.~\ref{fig:1}b).

A boundary condition 
at the heterostructure interface is a way to
establish the interface scattering properties with respect to an
incident wave\cite{OhkawaUemuraII1977,ShamNakayama1979}
with a wave vector $\bm{k}$
close to the band minima.
At the Si heterostructure, due to $z$-confinement,
there appear a mixing\cite{Ohkawa1978SolStatComm} between
the two low-energy valley states\cite{Boykin2004,Nestoklon2-2006,Jancu2005,Chutia2007}
at $\bm{k}_0$ and $-\bm{k}_0$ (Fig.~\ref{fig:1}a and b),
%
%
which implies  intra-valley or inter-valley scattering.
%
Generally, the
scattering off the interface
may lead not only to intervalley tunneling transitions ($\bm{k}_0 \to -\bm{k}_0$),
but also to a
spin-flipping\cite{GolubIvchenko2004,Boykin2004,Nestoklon2-2006,Jancu2005,Nestoklon3-2008,Rodina1Alekseev2006},
$\sigma \rightarrow -\sigma$ (see below).

Assuming the generalized envelope functions \cite{KohnLuttinger},
the total electron wave function is written
in the single-band approximation
 as:
\begin{equation}
%
\Psi(\bm{r}) = \left[\Phi_{\hat{z}}(\bm{r}) \psi_{\bm{k}_0}(\bm{r}) + \Phi_{-\hat{z}}(\bm{r}) \psi_{-\bm{k}_0}(\bm{r}) \right]
\label{KL_wave_function}
\end{equation}
where
the Bloch functions at the two band minima (at the $\Delta$ points) are
$\psi_{\pm\bm{k}_0}(\bm{r}) = e^{\pm i \bm{k}_0 z} u_{\pm\bm{k}_0}(\bm{r})$,
and
$u_{\pm\bm{k}_0}(\bm{r})$ are the periodic amplitudes.
The $\Phi_{\pm\hat{z}}(\bm{r})$ are spinor envelopes corresponding to the two valleys:
$\Phi_{\hat{z}}(\bm{r}) = [\Phi_{\hat{z},\uparrow}(\bm{r}),\Phi_{\hat{z},\downarrow}(\bm{r})]^T$ and
$\Phi_{-\hat{z}}(\bm{r}) = [\Phi_{-\hat{z},\uparrow}(\bm{r}),\Phi_{-\hat{z},\downarrow}(\bm{r})]^T$,
with spin components $\sigma = \uparrow,\downarrow$;
the envelopes $\Phi_{\pm\hat{z}}(\bm{r})= \Phi_{x,y}(x,y)\, \Phi_{\pm\hat{z}}(z)$
are separable in the absence of magnetic field.

In what follows, we consider an equivalent representation,
in which the state is described as a four-component
vector
\begin{equation}
\Phi(\bm{r}) \equiv
[\Phi_{\hat{z},\uparrow}(\bm{r}),\Phi_{\hat{z},\downarrow}(\bm{r}),
\Phi_{-\hat{z},\uparrow}(\bm{r}),\Phi_{-\hat{z},\downarrow}(\bm{r}) ]^T
\label{4-vector},
\end{equation}
subject to  boundary conditions and tunneling Hamiltonians.

\subsection{Boundary conditions for  $\rm Si/SiO_2$  heterostructure}
\label{Boundary_conditions}

The effective boundary condition  
at the $\rm Si/SiO_2$-interface
will act on the four-component   
envelope
$\Phi(\bm{r})$, Eq.~(\ref{4-vector}),
and it is derived  from symmetry reasonings,
for an infinitely high barrier (assuming a left interface at $z=z_0^{+} \equiv z_0 + \varepsilon,\ \varepsilon \to +0$):
\begin{equation}
\left\{ 1 + i R\, k_z - R\, \frac{2m_l}{\hbar^2}\, V_{\rm if}(\bm{k}) \right\} \Phi(\bm{r})\mid_{z=z_0^{+}}
\equiv {\cal B} \Phi(\bm{r})\mid_{z=z_0^{+}} = 0
\label{BC}  .
\end{equation}
%
Here $k_j \equiv -i \partial_j$ are quasi-momentum operators ($j=x,y,z$),
${\cal B}$ is a boundary operator,
$R$ is a parameter of dimension of length, characterizing
an abrupt interface\cite{VolkovPinsker1979,Vasko1979},  
and it is assumed that $R \ll l_z, l_D$, where $l_z, l_D$ are the QD confinement
lengths
along $z$-direction and in lateral directions.
%
For $R=0$,  Eq.(\ref{BC}) reduces to the standard BC, $\Phi(z)\mid_{z=z_0^{+}} = 0$
(which is unphysical, see Appendix \ref{App C: R-parameter}).
For $R\neq 0$  the BC leads to spin and valley mixing at the interface
via the  $4\times 4$ mixing matrix $V_{\rm if}(\bm{k})$ described in the next Sec. \ref{sec: interface-mixing-C2v}.

The form of the BC, Eq.(\ref{BC}),
can be  understood through the
general transfer matrix formalism\cite{AndoMori82},
where hermiticity of the Hamiltonian across the interface is preserved
using a transfer matrix $\hat{T}$ (has to be hermitean either) that relates the envelope function and its
derivative normal to the interface on both sides of the interface
(see also Ref.\cite{Tokatly2002,Rodina1Alekseev2006} for a recent account).
E.g., for the left interface for a single band and in the case of infinitely high barrier
(spin-valley mixing is dropped for a while):
\begin{equation}
0 = \left[
\begin{array}{c}
 \Phi(z_0^{-})\\
 \partial_z \Phi(z_0^{-})
\end{array}
\right] =
\left(
\begin{array}{cc}
 T_{11}& T_{12}\\
 T_{21}& T_{22}
\end{array}
\right)
\left[
\begin{array}{c}
 \Phi(z_0^{+})\\
 \partial_z \Phi(z_0^{+})
\end{array}
\right]
\label{AndoMoriBC} ,
\end{equation}
and a non-trivial solution of (\ref{AndoMoriBC}) implies the ``resonant condition''\cite{Braginsky1998}
${\rm det} \hat{T} = 0$; so, $T_{12} \neq 0$.
This means the relation
\begin{equation}
\Phi(z_0^{+}) + i R\, k_z \Phi(z_0^{+}) = 0
\label{VolkovPinskerBC1} ,
\end{equation}
reproducing the first two terms in (\ref{BC})
with $R \equiv T_{12}/T_{11}$, and
implying a discontinuity of the wave function and its derivative at the interface:
$\Phi(z_0^{+})\neq 0$ and $k_z \Phi(z_0^{+}) \neq 0$.
In the last form, using the dimensional interface parameter $R$,
the BC was first derived in Ref.\onlinecite{VolkovPinsker1979},
by requiring preservation of the hermiticity of
the Hamiltonian in the half-space,
$z>z_0$.
Physically, this implies continuity of the envelope flux density \cite{Rodina1Alekseev2006,VolkovPinsker1979}
(see also Appendix \ref{App C: Volkov-Pinsker}).
The parameter $R$, as well as the transfer matrix $\hat{T}$, is
a characteristics of the interface boundary region; here, we will take it as a phenomenological parameter.
An estimation,
based on a two-band model
(Appendix \ref{App C: R-parameter})
gives $|R| \approx 0.1 - 0.2\, {\rm nm}$ in the case of a $\rm Si/SiO_2$-interface.

If one drops the $k_z$-term in Eq.~(\ref{BC}), then the BC is of the usual
``non-resonant type'' (in the sense of Ref.\onlinecite{Braginsky1998}),
with $T_{12} = 0$,
and a transfer matrix obeys ${\rm det} \hat{T}_{\rm non-res} \neq 0$;
this implies a continuous envelop function
at the interface \cite{RodinaEfrosAlekseev2003}.
Such  BC have been suggested in Refs.\onlinecite{GolubIvchenko2004,Nestoklon2-2006,Nestoklon3-2008}
for the case of a Si/SiGe interface, and their ``non-resonant'' character make
them different from ours, Eq.~(\ref{BC}).

In this paper we suggest that the surface contributions
associated with the $k_z$-term can be important.
In particular, the
interface contribution to the
$g$-factor change will be zero without this term.
We also note, that for $R > 0$, it is possible to consider the so called Tamm states\cite{Tamm32},
(see also Refs.\onlinecite{VolkovPinsker1979,Vasko1979,Rodina1Alekseev2006}),
leading to localization in the $\hat{z}$-direction even in the absence of electric field
(to be considered elsewhere).

\subsection{The $C_{\rm 2v}$ interface mixing matrix}
\label{sec: interface-mixing-C2v}

The spin-valley mixing interface matrix $V_{\rm if}(\bm{k})$ that enters the BC
(\ref{BC}), can be expressed by
taking into account the $C_{\rm 2v}$ symmetry   
at the $\rm Si/SiO_2$ interface
(see, e.g. Refs.\cite{RashbaSheka-book1991,GolubIvchenko2004,Nestoklon2-2006,Nestoklon3-2008}${}^{,}$
\footnote{For ideal quantum well interfaces the relevant
interface symmetry ($D_{\rm 2d}$ or $D_{\rm 2h}$) admits only the invariant structure
corresponding to a Dresselhaus contribution \cite{GolubIvchenko2004},
while 
with an applied perpendicular electric field the reduced $C_{\rm 2v}$ symmetry admits also the Rashba structure.}).
The relevant $C_{\rm 2v}$-invariants are
the Rashba and Dresselhaus $2\times 2$ forms:
$H_R(\bm{k}) = \sigma_x k_y - \sigma_y k_x$,  $H_D(\bm{k}) = \sigma_x k_x - \sigma_y k_y$.
Indeed,
for the $C_{\rm 2v}$-symmetry transformations\cite{KiselevIvchenkoRossler1998PRB,Tokatly2002} one gets:
(i) a $\pi_z$-rotation leading to $k_{x,y} \to - k_{x,y}$ and $\sigma_{x,y} \to - \sigma_{x,y}$,
(ii) a reflection about the plane $(1,1,0)$, so that
$k_{x} \leftrightarrow - k_{y}$ and $\sigma_{x} \leftrightarrow \sigma_{y}$,
and (iii)
a reflection about the plane $(1,\bar{1},0)$, with
the $k_{x} \leftrightarrow k_{y}$ and $\sigma_{x} \leftrightarrow - \sigma_{y}$;
it is then easy to see that $H_R(\bm{k})$ and $H_D(\bm{k})$ remain unchanged under these
transformations.
Thus,
the $4\times 4$ spin-valley mixing matrix is parameterized as
\begin{eqnarray}
&& V_{\rm if}(\bm{k}) =
\left(
\begin{array}{cc}
 A(\bm{k}) & V \bm{I}_2  + B(\bm{k})\\
 V^* \bm{I}_2 + B^{+}(\bm{k})& A(\bm{k})
\end{array}
\right)
\label{spin-valley-mixing}\\
&& A(\bm{k})\equiv s_D\, H_D(\bm{k}) + s_R\, H_R(\bm{k})
\label{intra-vall} \\
&& B(\bm{k})\equiv \chi_D\, H_D(\bm{k}) + \chi_R\, H_R(\bm{k})
\label{inter-vall} ,
\end{eqnarray}
where $s_{D,R}$ are real parameters,
while
the intervalley tunneling matrix elements
$V = |V| e^{i\phi_V(z_0)}$,
and $\chi_{D,R} = |\chi_{D,R}| e^{i\phi_{D,R}(z_0)}$ generally possess
phases \cite{Nestoklon2-2006,Nestoklon3-2008,Saraiva2009PRB}.
For a general choice of the origin 
the phases depend linearly on $z_0$,
$\phi_{V,D,R}(z_0) = \phi_{V,D,R} - 2 i k_0 z_0$,
as it follows from the original valley Bloch functions in Eq.~(\ref{KL_wave_function}).
The  $2\times 2$ block-diagonal element
$A(\bm{k})$
corresponds to {\it intra-valley} spin-flipping transitions.
The Rashba-type term  $s_R (\sigma_x k_y - \sigma_y k_x)$ in the BC  was previously
derived\cite{Vasko1979,Vasko-book1998} for single-valley semiconductors.
The constant $s_R$ has two contributions:
$s_R = s^{\rm bulk}_R + s^{\rm if}_R$ and it can be shown that
the bulk $g^*$-factor in Si can contribute to $s^{\rm bulk}_R$
(see, e.g. Refs.~\onlinecite{VaskoPrima1981,VaskoPrima1981}).
However,  in this paper we  argue
that interface contributions are dominating.
In particular,
at the interface, both Rashba and Dresselhaus contributions will be allowed.

The off-diagonal
elements $V \bm{I}_2$ and
$B(\bm{k})$
are related to an {\it inter-valley} tunneling    (in momentum space).
The non-spin-flipping  term ($\sim V$)  is responsible
for the valley splitting\cite{OhkawaUemuraI1977,OhkawaUemuraII1977,ShamNakayama1979}
(see also Refs.\cite{Saraiva2009PRB,Culcer2010PRB2,FriesenCoppersmith2010} for recent account).
The {\it inter-valley} spin-flipping process will be described by
the term $B(\bm{k})$.
One of the main results of this paper is the observation that just this {\it inter-valley} spin-flipping process
is dominating the description of the experimentally measured   
$g$-factor variations \cite{VeldhorstRuskov2015PRB}.

\subsection{Effective Hamiltonian for the $\rm Si/SiO_2$ heterostructure}

The effective two-valley Hamiltonian
acts on the four-component vector
$[\Phi_{\hat{z},\uparrow}(\bm{r}),\Phi_{\hat{z},\downarrow}(\bm{r}),
\Phi_{-\hat{z},\uparrow}(\bm{r}),\Phi_{-\hat{z},\downarrow}(\bm{r}) ]^T$
$\equiv\Phi(\bm{r})$,
and includes a bulk Si
(spin and valley  degenerate) part
\begin{equation}
{\cal H}_0 =  \left[ \sum_{j=x,y,z} \frac{\hbar^2 k_j^2}{2 m_j}  + U_{x,y} + U_z \right]\times \bm{I}_4
\label{bulk_H}
\end{equation}
with
the in-plane, $U_{x,y}$,
and perpendicular to the interface,
$U_z$,  confinement electron potentials
\begin{eqnarray}
&& U_{x,y} = \frac{m_t}{2} (\omega_x^2 x^2 + \omega_y^2 y^2)
\label{conf_potentials-xy} \\
&& U_z =    \left\{ \begin{array}{c}
                    |e| (z-z_0) F_z, \ \ z > z_0 \\
                    \infty , \ \ \ \ z < z_0
\end{array} \right.
\label{conf_potentials-z}  .
\end{eqnarray}
In what follows we consider a circular quantum dot \cite{FriesenCoppersmith2010},
$\omega_x = \omega_y \equiv \omega_0$, and assume a much stronger confinement in the $\hat{z}$-direction:
$l_z = (\hbar^2/2m_l |e| F_z)^{1/3} \ll l_D = (\hbar/m_t \omega_0)^{1/2}$,
where $m_l$, $m_t$ are the longitudinal and transverse effective masses
for $\Delta$-valley electrons, $|e|$ is the electron charge, and
$F_z$ is the $z$-confinement electric field.
For the parameters of the experiment
\cite{Yang2013NC,Veldhorst2014Nn,VeldhorstRuskov2015PRB},
for electric field $F_z \simeq 3\times 10^7\, {\rm V/m}$,
$l_z \approx 1\,{\rm nm}$.
The lateral QD size is
$l_D \approx 7\,{\rm nm}$ for the 1e-case: $\Delta^{1e}_{\rm orb}\equiv \hbar\omega_0 \simeq 8\, {\rm meV}$;
for the 3e-case, $l_D \approx 14\,{\rm nm}$: $\Delta^{3e}_{\rm orb} \simeq 2\, {\rm meV}$
(since the ``valence electron'' in this case ``sees'' Coulomb repulsion, Figs.~\ref{fig:1}c and d).
Here, $\Delta^{1e,3e}_{\rm orb}$, are the usual orbital splittings in the QD, Fig.~\ref{fig:1}d.

The BC (\ref{BC})
induces a $\delta$-functional Hamiltonian contribution, ${\cal H}_{\rm if}$
that mixes the spin and valley states:
%
\begin{eqnarray}
&&{\cal H}_{\rm if} = -\frac{\hbar^2}{2 R m_l} \delta(z-z_0) \mp i \frac{\hbar^2}{2 m_l} \delta(z-z_0) k_z
\nonumber \\
&& \qquad\qquad\qquad  {} + \delta(z-z_0) V_{\rm if}(\bm{k})
\label{interface_H} .
\end{eqnarray}
(To show Eq.~(\ref{interface_H}), one needs to integrate the Schr\"{o}dinger equation
with ${\cal H}_{\rm if}$ at the vicinity of the boundary, $z = z_0$.)
The $-$ ($+$) sign at the second term in Eq.(\ref{interface_H}) stands for left (right) interface,
with the replacement $z_0 = z_{\rm left}$ ($z_0 = z_{\rm right}$)
and, in general, the interface parameters at the two interfaces may be different, $R_{\rm left} \neq R_{\rm right}$).
For a strong enough electric field  the
$z$-confinement (Fig.~\ref{fig:1}b) will keep electrons  close to the
left interface ($l_z \ll d_{\rm QW} \equiv z_{\rm right} - z_{\rm left}$), and we will neglect
the influence of the right interface   
\footnote{Interference effects similar to that in Refs.~\cite{Nestoklon2-2006,Nestoklon3-2008}
will be considered elsewhere}.
We note that in the current experiment this is well fulfilled,
since the $\rm {}^{28}Si$ QW thickness is $d_{\rm QW} \approx 300 - 800\, {\rm nm}$, while $l_z \approx 1\, {\rm nm}$
for $F_z \simeq 3\times 10^7\, {\rm V/m}$.
%
%
Since $l_z \propto F_z^{-1/3}$,  smaller electric fields are possible, 
providing
the z-confinement energy splitting is
much larger than the orbital splitting:
$1.5\hbar^2/(m_l l^2_z) \gg \hbar^2/(m_t l^2_D)$; 
e.g., for $\Delta_{\rm orb} = 1\, {\rm meV}$ one gets a typical  
field of $F_z \simeq 1.3\times 10^6\, {\rm V/m}$.

\section{Valley splitting, 2D(3D) effective Hamiltonians, and interface symmetry}

\subsection{The effective interface perturbation Hamiltonian}

The interface contribution, Eq.(\ref{interface_H}), is essentially singular   
and cannot be used, in general, as a perturbation (except in a heuristic way).
The effective interface perturbation Hamiltonian can be obtained
by recasting the original problem of the Hamiltonian ${\cal H}_0$, Eq.(\ref{bulk_H}), plus
boundary conditions Eq.(\ref{BC}),
to a standard BC, ${\cal B}\Phi\mid_{z=z_0^{+}} \equiv \tilde{\Phi}\mid_{z=z_0^{+}}=0$, and a transformed Hamiltonian.
To this end we consider the 3rd term in the BC Eq.(\ref{BC}) as a perturbation
(as $\langle k^2_{x}\rangle, \langle k^2_{y}\rangle \ll \langle k^2_z \rangle$) and
replace the boundary operator ${\cal B}$  up to higher orders
with a suitable  unitary transform $\Gamma_{\rm BC}$ (Appendix \ref{App A: eff-Hamiltonian}):
\begin{eqnarray}
&& \tilde{\Phi}\mid_{z=z_0^{+}} \simeq \Gamma_{\rm BC}\Phi\mid_{z=z_0^{+}} = 0
\label{transformed-BC}\\
&& \tilde{{\cal H}} = \Gamma_{\rm BC} {\cal H}_0 \Gamma^{\dagger}_{\rm BC} \simeq
{\cal H}_0 + \delta {\cal H}
\label{transformed-H} ,
\end{eqnarray}
with
$\Gamma_{BC} = 1 + i [R k_z + R^2 \frac{2m_l}{\hbar^2} V_{\rm if}(\bm{k}) k_z]$.
Keeping only the leading contribution in (\ref{transformed-H}) of order ${\cal O}(R^2)$, one obtains:
\begin{equation}
\delta {\cal H}(z)  \simeq R \partial_z U_z + R^2 \frac{2m_l}{\hbar^2} V_{\rm if}(\bm{k}) \partial_z U_z
\label{eff-interface-Hamiltonian} .
\end{equation}
In the following  we will neglect the first term in Eq.(\ref{eff-interface-Hamiltonian}) which leads
to a common energy shift only.

\subsection{Approximate diagonalization of the interface matrix. Valley splitting}
\label{sec: valley-splitting}

As suggested by the experiment \cite{VeldhorstRuskov2015PRB}, the valley splitting
matrix element
is much stronger  than the corresponding  
spin matrix elements \cite{valley-vs-spin-splitting},
$|V| \gg \{|\chi_{R,D}|, s_{R,D}\} \langle k_{x,y}\rangle$,
and the interface spin-valley matrix is represented as
$V_{\rm if}(\bm{k}) = V_{\rm if,val} + {\cal O}(1/|V|)$
with
%
\begin{equation}
V_{\rm if,val}
 =
\left(
\begin{array}{cc}
0 & V \bm{I}_2 \\
V^* \bm{I}_2 & 0
\end{array}
\right) .
\label{valley-splitting}
\end{equation}
%
%
%
Thus, one diagonalizes the interface Hamiltonian, Eq.~(\ref{eff-interface-Hamiltonian}),
to leading order via the unitary transform
(we choose below $z_0=0$ for convenience)  
\begin{equation}
U_{\rm v} =
\frac{1}{\sqrt{2}}\left(
\begin{array}{cc}
 \bm{I}_2 & -e^{i\phi_V} \bm{I}_2 \\
 e^{-i\phi_V} \bm{I}_2 & \bm{I}_2
\end{array}
\right)
\label{Unitary-v} ,
\end{equation}
%
%
%
leading to spin-independent valley-splitting Hamiltonian
\begin{equation}
\delta {\cal H}_{\rm if,val} =
\frac{2m_l}{\hbar^2} R^2 \, V^{\rm d}_{\rm if,val}\, \partial_z U_z
\label{val-splitting-H},
\end{equation}
with $V^{\rm d}_{\rm if,val} = {\rm diag}(|V| \bm{I}_2, -|V| \bm{I}_2)$.
The
corresponding
spin-degenerate eigenstates  
are denoted as
$|v^{\rm d}_{2,\sigma}\rangle = [C^T_\sigma,0,0]^T$  and
$|v^{\rm d}_{1,\sigma}\rangle = [0,0,C^T_\sigma]^T$
for the upper and lower eigenvalley states, respectively;
$C_\sigma$ is a spinor, corresponding to the two spin projections along an applied $\bm{B}$-field.
Turning back to
the original $\pm\hat{z}$-valley  basis, the eigenstates
of the leading-order Hamiltonian ${\cal H}_0 + \delta{\cal H}_{\rm if,val}$  will be written as
\begin{eqnarray}
&& |\bar{v}_{i;\sigma}\rangle = \frac{1}{\sqrt{2}}
\left[
\begin{array}{c}
C_\sigma \\
\mp e^{-i\phi_V} C_\sigma
\end{array} \right]\, \phi_0(x,y) \, \tilde{\varphi}_0(z) ,\ \ i=1,2
\label{v1-v2-eigenstates}
\end{eqnarray}
where $\phi_0(x,y)\, \tilde{\varphi}_0(z)$ is an eigenstate of
${\cal H}_0$, Eq.(\ref{bulk_H}),
with BC, $\tilde{\varphi}_0(0^{+}) = 0$,
in the lowest $z$-subband.
%
The upper/lower  eigenvalley energies are
$E_{2,1} = \langle \bar{v}_{2,1} | \delta{\cal H}_{\rm if,val}   |\bar{v}_{2,1}\rangle =
\pm \frac{|V| 2 m_l  R^2}{\hbar^2}\, \langle \tilde{\varphi}_0(z) |\partial_z U_z|\tilde{\varphi}_0(z) \rangle
 \equiv \pm \frac{|V| 2 m_l  R^2}{\hbar^2}\, \langle \partial_z U_z \rangle$
and the valley splitting reads:
\begin{equation}
E_{\rm VS} = 2 |V| R^2 \frac{2 m_l}{\hbar^2}\, \langle \tilde{\varphi}_0(z) |\partial_z U_z|\tilde{\varphi}_0(z) \rangle
\label{valley_splitting} .
\end{equation}
By observing the general integral relation (Appendix \ref{App B: integral relation})
\begin{equation}
\langle \tilde{\varphi}(z) |\partial_z U_z|\tilde{\varphi}(z) \rangle
\equiv \int_0^{\infty}\, dz \tilde{\varphi}^*(z) \partial_z U_z \tilde{\varphi}(z)  = \frac{\hbar^2}{2 m_l}\, |\tilde{\varphi}^{\prime}(0)|^2
\label{integral_relation}
\end{equation}
[It holds for any eigenstate of the Hamiltonian (\ref{bulk_H})  with a smooth
(at $z > 0$)
$z$-confinement potential $U_z$
and zero BC, $\tilde{\varphi}(0) = 0$],
one can recast the valley splitting to the form
\begin{equation}
E_{\rm VS} =  2 |V| R^2 \, |\tilde{\varphi}_0^{\prime}(0)|^2
\label{valley_splitting1} .
\end{equation}

Alternatively, the valley splitting
can be derived
in a different (heuristic) way, using the singular Hamiltonian, Eq.(\ref{interface_H}).
In this case, one would consider the first two terms in Eq.(\ref{interface_H})
as a leading order  boundary condition, recasting them to
the Volkov-Pinsker form\cite{VolkovPinsker1979}
\begin{equation}
[1 + R\partial_z] \varphi_0(0) = 0
\label{BC_zeroth_order} ,
\end{equation}
[scf. Eq.(\ref{BC})].
Since R is small, one essentially has the BC $\varphi_0(R) = 0$
which corresponds to $z$-shifting the origin by R.
With $\varphi_0(z)$ being the eigenstate of the Hamiltonian (\ref{bulk_H}) ${\cal H}_0$
with the above BC (\ref{BC_zeroth_order})
one considers the ``perturbation'' $\delta(z)\, V^{\rm d}_{\rm if,val}$
from Eq.(\ref{interface_H}),
with the diagonal part of the
interface matrix.
This gives the valley splitting
\begin{equation}
E_{\rm VS} =  2 |V| |\varphi_0(0)|^2 \simeq 2 |V| R^2 |\varphi_0^{\prime}(0)|^2
\simeq 2 |V| R^2 |\tilde{\varphi}_0^{\prime}(0)|^2
\label{valley_splitting2} ,
\end{equation}
where we have used Eq.~(\ref{BC_zeroth_order}),
and that
$\tilde{\varphi}_0^{\prime}(0) \simeq {\varphi}_0^{\prime}(0)$ up to higher orders in R.
The result, Eq.(\ref{valley_splitting2}), for the valley splitting  coincides
with
Eqs.~(\ref{valley_splitting}) and (\ref{valley_splitting1}),
obtained via the effective Hamiltonian Eq.(\ref{eff-interface-Hamiltonian}).

Notice that for ${\cal H}_0$, Eq.(\ref{bulk_H}), with the linear $z$-confinement potential $U_z = |e| F_z z$
(the ``triangular'' potential)
one has 
the lowest energy subband  
function:
$\tilde{\varphi}_0(z) = N_1 \, l_z^{-1/2}\, {\rm Ai}(l_z^{-1} z - \tilde{E}_1)$
with a normalization
$N_1 \simeq 1.4261$, and
$-\tilde{E}_1 = -2.3381$  being the first zero of the Ai function.
The $z$-average is
$\langle z\rangle \simeq  1.5587\, l_z = 1.5587\, (\hbar^2/2m_l |e| F_z)^{1/3}$,
see Eq.(\ref{bulk_H}).
For the valley splitting one gets then from  Eq.(\ref{valley_splitting}):
\begin{equation}
E_{\rm VS} = 2 |V| R^2\, \frac{2 m_l |e| F_z}{\hbar^2} = 2 |V| R^2 l_z^{-3}
\label{valley_splitting3}  .
\end{equation}
Thus, the  general
relation Eq.~(\ref{valley_splitting1}) we have proven, (Appendix \ref{App B: integral relation}) is fulfilled
here from the relation
$\frac{d \tilde{\varphi}_0(z)}{dz} = N_1 \, l_z^{-3/2}\, {\rm Ai}^{\prime}(l_z^{-1} z - \tilde{E}_1)$
and by noticing that
$N_1 {\rm Ai}^{\prime}(-\tilde{E}_1) = 1$.

For the second (heuristic) approach, with the ``shifted BC'' Eq.(\ref{BC_zeroth_order}),
the eigenstates of the Hamiltonian, Eq.(\ref{bulk_H})
will be just the shifted functions, with the lowest subband being:
\begin{equation}
\varphi_0(z) = N_1 \, l_z^{-1/2}\, {\rm Ai}(l_z^{-1} (z-R) - \tilde{E}_1) ,
\label{shifted-wave-function}
\end{equation}
and $|\varphi_0(0)| = R |\varphi^{\prime}_0(0)| \simeq R |\tilde{\varphi}^{\prime}_0(0)| \neq 0$,
as implied by Eq.(\ref{valley_splitting1}) and the Volkov-Pinsker BC, Eq.(\ref{BC_zeroth_order}).

The linear dependence  on $F_z$, Eq.~(\ref{valley_splitting3}), is confirmed experimentally\cite{Yang2013NC,Veldhorst2014Nn}.
%
Using the estimation $R \approx 0.1\, {\rm nm}$ (Appendix \ref{App C: R-parameter})
and the experimental slope \cite{Yang2013NC}
$\frac{\Delta E_{\rm VS}}{\Delta F_z} = 1.32\,\, {\rm e\,{\AA}}$
one gets a valley-splitting parameter $|V| \approx 2640\,\, {\rm meV}\, {\rm {\AA}}$
compatible with the effective mass and tight-binding calculations \cite{Chutia2007,Nestoklon3-2008}
(extrapolated to the $\rm Si/SiO_2$ case\cite{V-simple-tight-binding}).

Eq.~(\ref{valley_splitting3})
 corresponds to a valley splitting with linear $F_z$-dependence  
 and no offset,
applicable for relatively large electric fields, $F_z \gtrsim 3\times 10^7 {\rm V/m}$,
when $z$-confinement is much stronger than lateral confinement.
(Notice, however, that for larger QDs our results are applicable
at lower electric fields as well).
On the other hand, the measurements of the valley splitting in our previous work \cite{Yang2013NC,Veldhorst2014Nn}
suggest that such offset could be possible.
For example, a possible non-linear dependence at small electric field suggested
by tight-binding calculations \cite{Boykin2004,Nestoklon3-2008}
could lead to an effective offset.

Here we propose   
a phenomenological approach that allows   
to describe the experimentally observed valley splitting offset\cite{Yang2013NC,Veldhorst2014Nn}
resulting from an interface localized interaction.
Using the general results, Eqs.~(\ref{valley_splitting})-(\ref{valley_splitting2}),
one considers a confinement potential of the form
$U_z = \frac{1}{2}m_l\omega_z^2 z^2 + |e| z F_z$, which provides a non-zero valley splitting at $F_z=0$,
with a confinement length factor,
$l_{\rm osc}^{-3} \equiv |\tilde{\varphi}_0^{\prime}(0)|^2 = \frac{8}{\sqrt{\pi}} \left(\frac{m_l\omega_z}{\hbar}\right)^{3/2}$.
In the opposite limit of large $F_z$, the zero-field confinement can be considered as a perturbation
to the linear potential, leading asymptotically to the behavior:
$E^{\rm asympt}_{\rm VS} \simeq 2 |V| R^2\, \frac{2 m_l }{\hbar^2} ( |e| F_z + m_l \omega_z^2 \langle z \rangle )$,
which can be interpreted as a positive offset.
To obtain a negative offset,
one needs to replace the interface-localized confinement with a repulsion $z$-potential.

\subsection{Approximate diagonalization of the interface matrix:
The 2D Spin-Orbit Dresselhaus and Rashba couplings and effective 2D (3D) Hamiltonians}

The effective 
spin-orbit Hamiltonians (of Rashba and Dresselhaus type)
are obtained similarly to the $E_{VS}$ calculation.
For this end, we apply now
the unitary transformation $U_{\rm v}$, Eq.~(\ref{Unitary-v}),
to the full interface matrix,
$V^{\rm U}_{\rm if}(\bm{k})= U_{\rm v}^{+} V_{\rm if}(\bm{k})U_{\rm v}$
and obtain the form:
\begin{eqnarray}
&& V^{\rm U}_{\rm if}(\bm{k}) =
\left(\!\!
\begin{array}{cc}
 |V| \bm{I}_2  &  0 \\
 0 & -|V| \bm{I}_2
\end{array}
\!\!\right)
+
\left(\!\!
\begin{array}{cc}
  A + \frac{1}{2} B_{\rm diag}  &  \frac{1}{2} B_{\rm off} \\
 {\rm h.c.} & A - \frac{1}{2} B_{\rm diag}
\end{array}
\!\!\right) \quad \ \
\label{spin-valley-mixing-unitary-transformed}\\
&& \qquad\quad  \equiv V^{\rm d}_{\rm if,val} +
V^{\rm s-v}_{\rm if}(\bm{k}) ,
\nonumber
\end{eqnarray}
with
\begin{eqnarray}
&& B_{\rm diag} \equiv B(\bm{k}) e^{-i\phi_V(z_0)}+ B^{+}(\bm{k}) e^{i\phi_V(z_0)},
\label{B-diag}\\
&& B_{\rm off} \equiv B(\bm{k}) - B^{+}(\bm{k}) e^{2i\phi_V(z_0)},
\label{B-off}
\end{eqnarray}
obtained via Eq.(\ref{inter-vall}), with $\phi_V(z_0) = \phi_V - 2 i k_0 z_0$.

The  
spin-valley mixing part in (\ref{spin-valley-mixing-unitary-transformed}),
$V^{\rm s-v}_{\rm if}(\bm{k})$,
consists of the (eigen)valley block-diagonal  and off-diagonal parts
and constitutes the spin-orbit effective coupling at the interface,
derived from Eq.~(\ref{eff-interface-Hamiltonian}):
\begin{eqnarray}
&&
\delta {\cal H}_{\rm s-v} = R^2 \frac{2m_l}{\hbar^2}\, V^{\rm s-v}_{\rm if}(\bm{k})\,\partial_z U_z
\nonumber\\
&& \qquad\quad { } =
V^{\rm s-v}_{\rm if}(\bm{k})\, R^2 |\varphi_0^{\prime}(0)|^2 \frac{\partial_z U_z}{\langle \partial_z U_z \rangle}
\label{spin-val-H}  ,
\end{eqnarray}
with matrix elements between the eigenvalley states $v_1$, $v_2$,
that are proportional to the Rashba and Dresselhaus invariant forms, $H_R(\bm{k})$, $H_D(\bm{k})$.
The spin-valley mixing Hamiltonian $\delta {\cal H}_{\rm s-v}$, Eq.~(\ref{spin-val-H}),
then reads:
\begin{eqnarray}
&&
\delta {\cal H}_{\rm s-v} =
\nonumber\\
&&
{ } = \left(\!\!\!\!
\begin{array}{cc}
 \alpha_{R;v_2} H_R + \beta_{D;v_2} H_D, &\!\!  \alpha_{R;21} H_R + \beta_{D;21} H_D \\
 \alpha_{R;21}^{*} H_R + \beta_{D;21}^{*} H_D, &\!\! \alpha_{R;v_1} H_R + \beta_{D;v_1} H_D
\end{array}
\!\!\!\!\right)\!\! \frac{\partial_z U_z}{\langle \partial_z U_z \rangle} ,\ \ \ \
\label{spin-valley-mixing-3D}
\end{eqnarray}
%
where $\alpha_{R;v_i}$, $\beta_{D;v_i}$, and
$\alpha_{R;21}$, $\beta_{D;21}$ are the diagonal and off-diagonal
(valley dependent) Rashba and Dresselhaus  coupling constants,
related to the effective SOC interactions considered below.
We derive the SOC constants, taking into account the phases
of
$\chi_R = |\chi_R| e^{i\phi_R(z_0)}$, $\chi_D = |\chi_D| e^{i\phi_D(z_0)}$
in a translationally invariant form \cite{translation-invariance-0}.
For the diagonal constants one obtains:
\begin{eqnarray}
&& \alpha_{R;v_i} = \left[ s_R \mp |\chi_R| \cos(\phi_R-\phi_V)\right] \, R^2 |\varphi_0^{\prime}(0)|^2
\label{SOC-consts-a11}\\
&& \beta_{D;v_i} = \left[ s_D \mp |\chi_D| \cos(\phi_D-\phi_V)\right] \, R^2 |\varphi_0^{\prime}(0)|^2
\label{SOC-consts-b11}\\
&& i=1,2
\nonumber
\end{eqnarray}
with
$-$ ($+$) corresponding to the lower eigenvalley $v_1$ (upper eigenvalley $v_2$) respectively;
this is similar to the relevant strong field limit results of Ref.\onlinecite{Nestoklon3-2008}.
%
The off-diagonal
Rashba and Dresselhaus coupling constants are,
correspondingly:
\begin{eqnarray}
&& \alpha_{R;21} = i e^{i \phi_V} |\chi_R| \sin(\phi_R-\phi_V)\, R^2 |\varphi_0^{\prime}(0)|^2
\label{SOC-consts-a21}\\
&& \beta_{D;21} = i e^{i \phi_V} |\chi_D| \sin(\phi_D-\phi_V)\, R^2 |\varphi_0^{\prime}(0)|^2
\label{SOC-consts-b21} .
\end{eqnarray}
Notice that for a linear $z$-confinement,
Eq.~(\ref{conf_potentials-z}),  
the SOC constants scale linearly with the applied electric field $F_z$.
%
The off-diagonal elements $\alpha_{R;21}$, $\beta_{D;21}$
could be,
generally, of the same order as the diagonal one, $\alpha_{R;v_i}$, $\beta_{D;v_i}$,
depending on the phases, $\phi_V$, $\phi_R$, $\phi_D$, and assuming $|\chi_{R,D}| \gtrsim s_{R,D}$.
These parameters, including the phases, enter in the observable SOC constants
in certain combinations, relating the diagonal to off-diagonal (in valley)
SOC constants, Eqs.~(\ref{SOC-consts-a11})-(\ref{SOC-consts-b21}).
%
Eq.~(\ref{spin-valley-mixing-3D}) and Eqs.~(\ref{SOC-consts-a11})-(\ref{SOC-consts-b21})
describe the 3D spin-valley mixing at the interface.
These equations are one of the main results of this
paper, together with the $g$-factor derivation in the next chapter, which will be
based on them as well.

A 2D version can be obtained by integration over the $z$-direction.
The effective 2D Hamiltonian  with  Rashba and Dresselhaus contributions in each eigenvalley subspace
is given by the corresponding block-diagonal parts in Eq.~(\ref{spin-valley-mixing-3D}):
\begin{equation}
{\cal H}^{\rm 2D}_{v_i} = \alpha_{R;v_i}\, H_R(\bm{k}) + \beta_{D;v_i}\, H_D(\bm{k}),\ \ i=1,2
\label{2D-Hamiltonian-diag} ,
\end{equation}
with
the 2D spin-orbit couplings given by
Eqs.~(\ref{SOC-consts-a11})  and (\ref{SOC-consts-b11}).
%
Similarly, the 2D Hamiltonian that describes the off-diagonal transitions
between the eigenvalley subspaces $v_1,v_2$ can be written in the form
\begin{equation}
{\cal H}^{\rm 2D}_{v_2,v_1} = \alpha_{R;21} H_R(\bm{k}) + \beta_{D;21} H_D(\bm{k})
\label{2D-Hamiltonian-non-diag} ,
\end{equation}
with the 2D spin-orbit couplings given by
Eqs.~(\ref{SOC-consts-a21})  and (\ref{SOC-consts-b21}).

As seen from Eqs.~(\ref{SOC-consts-a11})-(\ref{SOC-consts-b21}),
all the above
spin-orbit constants
depend
on the common matrix elements constants, $V$, $s_R$, $s_D$, $\chi_R$, $\chi_D$, that parameterize
the  spin-valley mixing boundary condition, Eq.~(\ref{BC}).
We note, that the 2D spin-orbit Rashba and Dresselhaus constants, $\alpha_{R;v_i}$, $\beta_{D;v_i}$,
may change sign when one switches between the
eigenvalley subspaces $v_1 \to v_2$:
\begin{equation}
\alpha_{R;v_1} \simeq - \alpha_{R;v_2}\ {\rm and}\ \ \beta_{D;v_1} \simeq - \beta_{D;v_2}
\label{sign-SOC}
\end{equation}
if the inter-valley contributions, $\chi_R$, $\chi_D$ dominate the intravalley
ones, $s_R$, $s_D$;
Eq.~(\ref{sign-SOC}) is exact for $s_R,s_D = 0$.
As shown in the next Sec. IV, this is in
qualitative
agreement with the experiment \cite{VeldhorstRuskov2015PRB}, where
measurement of the $g$-factor were performed for an in-plane magnetic field.

Finally, we mention
that one can derive the 2D Hamiltonian Eq.~(\ref{spin-valley-mixing-3D})
without recasting the BC to a smooth perturbation Hamiltonian
[as it was done in Eqs.(\ref{transformed-H})  and (\ref{eff-interface-Hamiltonian})].
As in the valley splitting derivation in Eq.(\ref{valley_splitting2}),
one just refers to the leading order BC, Eq.(\ref{BC_zeroth_order}),
and uses (heuristically) the singular ``perturbation''
$\delta(z)\, V^{\rm U}_{\rm if}(\bm{k})$ with the full interface
matrix, Eq.(\ref{spin-valley-mixing-unitary-transformed}).
The effective interface Hamiltonian, Eq.~(\ref{eff-interface-Hamiltonian}), is
necessary, however,  for the derivation of the $g$-factor  where the heuristic approach
does not work.

\section{Electron $g$-factor at the interface}

\subsection{Derivation of the $g$-factor corrections}
\label{sec: Sec IV: g-factor correction}

We will consider for each eigenvalley the Hamiltonians,
Eqs.(\ref{bulk_H})  and (\ref{val-splitting-H}),
${\widetilde{\cal H}}_0 = {\cal H}_0 + \delta {\cal H}_{\rm if,val}$ as the zeroth-order term, and
%
the spin-valley mixing term  $\delta{\cal H}_{\rm s-v}$,
Eqs.~(\ref{spin-val-H})  and (\ref{spin-valley-mixing-3D}),
as a perturbation.
Since the valley splitting is large, one can neglect
the block-off-diagonal part in $\delta{\cal H}_{\rm s-v}$
as it contributes to the energy renormalization of the subspaces
$v_1$, $v_2$,  only in second order of PT,  and is suppressed as
$\sim |\chi_{D,R} \langle k_{x,y}\rangle|/E_{\rm VS}$.
%
The block-diagonal parts of $\delta{\cal H}_{\rm s-v}$
are of the form
\begin{equation}
{\cal H}_{v_i}^{\rm 3D} =
\left[ \alpha_{R;v_i}\, H_R(\bm{k}) + \beta_{D;v_i}\, H_D(\bm{k}) \right]\, \frac{\partial_z U_z}{\langle\partial_z U_z \rangle},\ \ i=1,2
\label{H-vi}  .
\end{equation}
One can note that these Hamiltonians are in one-to-one correspondence,
via Eqs.~(\ref{transformed-H})-(\ref{eff-interface-Hamiltonian}),
to the    
BCs in each eigenvalley subspace \cite{single-valley-BCs}:
\begin{equation}
\left\{ 1 + i R k_z - R \frac{2 m_l}{\hbar^2}
\left[ \mp |V| + V_{v_i}(\bm{k}) \right] \right\} \Phi_{v_i}(\bm{r})\mid_{z=z_0^{+}} = 0
\label{BC-spinor}  ,
\end{equation}
with the spin-mixing matrix
$V_{v_i}(\bm{k}) \equiv A \mp \frac{1}{2}\, B_{\rm diag}$ defined in
Eqs.~(\ref{intra-vall}), (\ref{inter-vall}), and (\ref{B-diag}),
and
acting on the corresponding eigenvalley spinors,
$\Phi_{v_i},\, i=1,2$.
Eq.~(\ref{H-vi}) may contribute to first order of PT
to the g-factor in each
eigenvalley subspace.

For a magnetic field  a direct Zeeman term is added to the
zeroth-order Hamiltonian ${\widetilde{\cal H}}_0$:
\begin{equation}
{\cal H}_{\rm Z} = g^* \mu_B \frac{1}{2}\bm{\sigma} \bm{B}
\label{Zeeman}
\end{equation}
where    $\mu_{\rm B}$ is the Bohr magneton;
the bulk Si effective $g^*$-factor \cite{WilsonFeher1960BullAPS,Roth1960PR,Liu1962PR},
is $g^*_{\rm Si} \simeq 1.9983$ (at the interface).

The perturbation due to external magnetic field will arise via
the replacement \cite{KohnLuttinger} $k_j \to k_j + \frac{|e|}{\hbar} A_j(\bm{r})$
[$\bm{A}(\bm{r})$ is the vector-potential],
both in ${\cal H}_0$ and in the interface Hamiltonian
$\delta{\cal H}_{\rm s-v}$  or, equivalently, in the respective BCs,
Eqs.~(\ref{BC}), (\ref{interface_H}),  and (\ref{BC-spinor}),
which makes the problem gauge invariant  
[For a gauge-invariant BC without spin and valleys, see Appendix \ref{App C: Volkov-Pinsker};
for a discussion of gauge-invariance see Appendix \ref{App C: Gauge-invariance}].
Introducing the magnetic length, $l_B = (\hbar/|e|B)^{1/2}$, we require a stronger $z$-confinement,
$l_z \ll l_B$, which is fulfilled in the experiment for $B=1.4\, {\rm T}$,
as  $l_B(1.4\, {\rm T}) \simeq 22\, {\rm nm}$.

\subsection{g-factor for in-plane magnetic field, $\bm{B}_{\parallel}$}

\subsubsection{$\delta g_{\parallel}$ to 1st-order PT}
\label{sec: g-parallel}

\begin{figure}[!h] 
    \centering
        \includegraphics[width=0.49\textwidth]{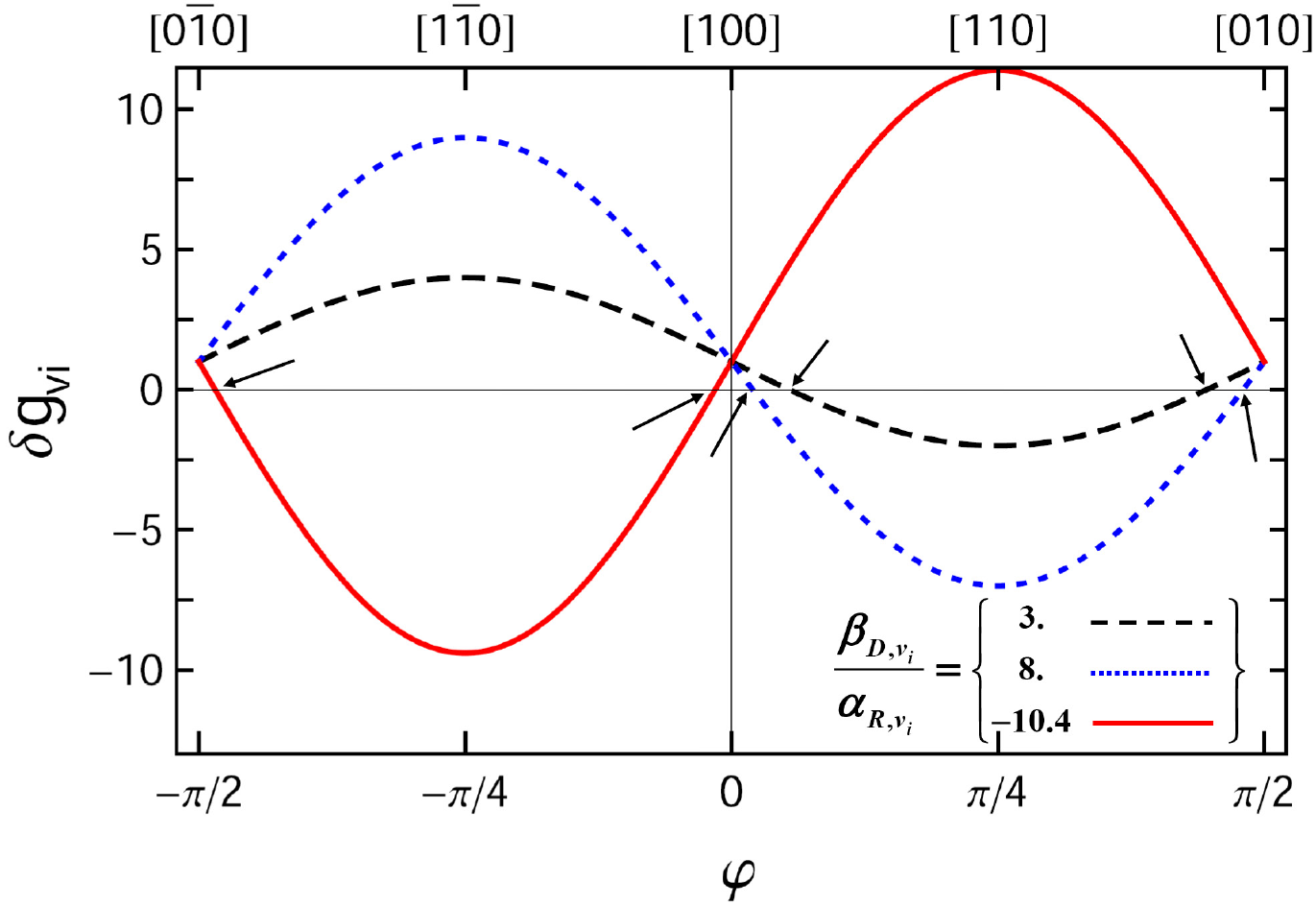}
        \caption{ Angular dependence
        of the $g$-factor correction $\delta g_{\parallel}^{v_i}(\varphi,F_z)$,
        Eq.~(\ref{g-eff-approx}) at different
        ratios of the
        spin-orbit parameters:
        $\beta_{D;v_i} / \alpha_{R;v_i} = \{3.0,\, 8.0,\, -10.4 \}$ (dashed black, dotted blue, and red curves, respectively);
        $\delta g_{\parallel}^{v_i}$ is in units of $\frac{e}{\hbar \mu_B} |\alpha_{R;v_i}|$.
        For $\bm{B}_{\parallel}$ at angles
        (shown with arrows)
        $\varphi_{v_i}$, Eq.~(\ref{optimal-angle-approx}), measured from the $[1,0,0]$
        crystal direction,  
        the QD spin-qubit is immune to the charge noise on $F_z$, since
        the g-factor variation due to electric field 
        noise 
        goes to zero together with $\delta g_{\parallel}^{v_i}(F_z)$,
        see Eqs.~(\ref{g-factor-derivative}), (\ref{g-factor-scaling}), and (\ref{new_charge_noise}).
         }
        \label{fig:2}
\end{figure}

For an in-plane magnetic field one chooses  the gauge $\bm{A}_{\|}(\bm{r}) = (B_y z, -B_x z, 0)$.
In what follows, we neglect small corrections originating from the bulk Hamiltonian ${\cal H}_0$, Eq.~(\ref{bulk_H}).
The perturbation to Eq.~(\ref{H-vi}), $\delta_B {\cal H}_{v_i}^{\rm 3D}$,
due to non-zero magnetic field $\bm{B}_{\parallel}$, contributes to
the g-factor interface contribution,
$\delta g^{v_i}_{\parallel}$,
to first order.
Averaging  
Eq.~(\ref{H-vi}) over the states   
$|\bar{v}_i\rangle \equiv |v_i\rangle \otimes |\phi_{v_i}(\bm{r})\rangle$
[that includes the envelope wave function of the confined electron
$|\phi_{v_i}(\bm{r})\rangle \equiv \phi_0^{v_i}(x,y) \varphi_0(z)$, see below],
for each eigenvalley
gives
\begin{eqnarray}
&& \langle \bar{v}_i | \delta_B {\cal H}_{v_i}^{\rm 3D} |\bar{v}_i\rangle = a[U_z]\ \mu_B \Big[ \alpha_{R;v_i} (\sigma_x B_x + \sigma_y B_y) \Big.
\nonumber\\
&& \qquad\qquad\qquad\ \  \Big. { } - \beta_{D;v_i} (\sigma_x B_y + \sigma_y B_x) \Big],\ \ i=1,2
\label{H_B_parallel}\\
&& \qquad  a[U_z] \equiv - \frac{|e|}{\hbar \mu_B} \frac{\langle z \partial_z U_z \rangle}{\langle \partial_z U_z \rangle}, \ \ \  a[U_z]\sim 10^{-3}
\label{a_general} ,
\end{eqnarray}
with the constant $a[U_z]$ being a weakly-dependent functional of the
z-confinement potential $U_z$.
For a constant electric field  $a[U_z]$ is replaced by
$- \frac{|e| \langle z \rangle}{\hbar \mu_B}$.
The total Zeeman energy can be written via the g-factor tensor:
\begin{equation}
{\cal H}^{\rm tot}_{\rm Z,v_i} =
\sum_{\alpha,\beta} \frac{1}{2}
\mu_B \left( g_0 \delta_{\alpha\beta} + \delta g^{v_i}_{\alpha\beta}\right) \sigma_{\alpha} B_{\beta}
\label{total_Zeeman} ,
\end{equation}
where
$g_0 = g^*_{\rm Si}$ is the bulk value in Si, and
\begin{eqnarray}
&& \delta g_{xx}^{v_i} = \delta g_{yy}^{v_i} = - a[U_z]\ \alpha_{R;v_i} 
\label{delta_g_long}\\
&& \delta g_{xy}^{v_i} = \delta g_{yx}^{v_i} = + a[U_z]\ \beta_{D;v_i}\, .  
\label{delta_g_trans}
\end{eqnarray}
The Zeeman splitting
is expressed as
$\Delta E \equiv \mu_B g_{\parallel}(\varphi) B_{\parallel}$,
$B_{\parallel} = \sqrt{B_x^2 + B_y^2}$,
and  $B_x = B_{\parallel} \cos \varphi$, $B_y = B_{\parallel} \sin \varphi$,  being the magnetic field components
along the Si crystal axes.
%
By  
diagonalization of the Hamiltonian (\ref{total_Zeeman})
for each valley subspace,
one obtains the  total g-factor  
$g_{\|}^{v_i}(\varphi,F_z)$,
\begin{eqnarray}
&& g_{\parallel}^{v_i}(\varphi,F_z) =
\Big( g_0^2 + 2a[U_z]\, g_0\,[\alpha_{R;v_i} - \beta_{D;v_i} \sin 2\varphi] + \Big. \qquad\qquad\quad
\nonumber\\
&& \qquad\quad  \Big. {} +
a[U_z]^2\,\, [\alpha^2_{R;v_i} + \beta^2_{D;v_i} - 2 \alpha_{R;v_i} \beta_{D;v_i} \sin 2\varphi] \Big)^{1/2}\!\!\!\! ,
\label{g_eff}
\end{eqnarray}
that includes
the interface contribution $\delta g_{\parallel}^{v_i}$:
$g_{\|}^{v_i} \equiv g_0 + \delta g_{\parallel}^{v_i}$.
The $F_z$ dependence in Eq.~(\ref{g_eff}) is implicit via the SOC constants and $z$-averages,
Eqs.~(\ref{SOC-consts-a11}),(\ref{SOC-consts-b11}), (\ref{integral_relation}), and (\ref{a_general}).
To first order in $a[U_z]$   
it gives the g-factor interface variation as a function of the in-plane magnetic  field
direction\cite{Delft2016-Ruskov-talk,QCPR2015-SanDiego-Tahan}, $\varphi$  (Fig.~\ref{fig:2}):
\begin{equation}
\delta g_{\parallel}^{v_i}(\varphi,F_z) \simeq
\delta^{(1)} g_{v_i} =
- \frac{|e|}{\hbar \mu_B} \frac{\langle z \partial_z U_z \rangle}{\langle \partial_z U_z \rangle}\,
\left( \alpha_{R;v_i} - \beta_{D;v_i} \sin 2\varphi \right)
\label{g-eff-approx} .
\end{equation}

The angular dependence on the direction of the in-plane magnetic  field suggests
that there could be valley-dependent  ``sweet spot directions''
$\varphi_{v_i}$
where the $g$-factor variation with the electric field  is  zero.
Since
from Eq.~(\ref{g-eff-approx}),
\begin{equation}
\frac{\partial[\delta g_{\parallel}^{v_i}(\varphi,F_z)]}{\partial F_z}
= \frac{\partial\ln \langle z\, \partial_z U_z \rangle}{\partial F_z} \ \delta g_{\parallel}^{v_i}(\varphi,F_z)
\label{g-factor-derivative}
\end{equation}
the $g$-factor noise variation   
gets to zero together with $\delta g_{\parallel}^{v_i}(F_z)$.
For a given eigenvalley $v_i$ the choice of the angle $\varphi_{v_i}$ will depend on the
size and sign of the Rashba and Dresselhaus 2D spin-orbit constants,
$\alpha_{R;v_i}$, $\beta_{D;v_i}$.
The 1st-order PT $g$-factor correction, Eq.~(\ref{g-eff-approx}), can be put to zero
when $\sin (2\varphi) = \alpha_{R;v_i}/\beta_{D;v_i}$.
Thus, the optimal angles are expressed as (Fig.~\ref{fig:2}):
\begin{equation}
\varphi_{v_i} = \frac{1}{2}\, \arcsin \left(\frac{\alpha_{R;v_i} }{\beta_{D;v_i}}\right)
\simeq \frac{\alpha_{R;v_i} }{2\beta_{D;v_i}} + k \frac{\pi}{2},\ \ k=0,\pm 1, \pm 2, \dots
\label{optimal-angle-approx} ,
\end{equation}
where the  inequality $|\alpha_{R;v_i}| \ll |\beta_{D;v_i}|$
is assumed from
tight binding calculations \cite{Nestoklon3-2008},
\footnote{R. Ferdous, R. Rahman, {\it private communication}}.
The sweet spot angles are
generally different for the two eigenvalley states $v_i$.
At these angles the spin qubit is immune to the charge noise
(via the electric field $F_z$, see Sec. \ref{sec: new_noise_mechanism}).
However, at the same sweet spot
angles the qubit frequency cannot be manipulated as well.
(From a qubit perspective, there should be a trade off, where one can keep the possibility
to manipulate the qubit  reasonably fast,
and simultaneously minimize the noise).  
There are weak second order PT effects, to be considered in the next section.
It is interesting to note that for a zero Dresselhaus contribution the
$g$-factor variation
$\delta g_{\parallel}^{v_i}$
becomes angle-independent.

For a linear $z$-confinement one can rewrite Eq.~(\ref{g-eff-approx})
as
\begin{equation}
\delta g_{\parallel}^{v_i}(\varphi,F_z) \equiv  A_{v_i}(\varphi) F_z^{2/3} ,
\label{g-factor-scaling}
\end{equation}
since
the SOC constants $\alpha_{R;v_i}, \beta_{D;v_i} \propto F_z$, and
the average of the $z$-motion in the lowest subband
is
$\langle z\rangle \simeq  1.5587\, (\hbar^2/2m_l |e| F_z)^{1/3}$,
see Eq.(\ref{bulk_H}).
In the 
experiment \cite{VeldhorstRuskov2015PRB},
where the magnetic field is parallel to
the $[110]$-direction (i.e. $\varphi = \pi/4$),
one gets from Eq.~(\ref{g-eff-approx}):
%
\begin{equation}
\delta g_{\parallel}^{v_i}(\pi/4,F_z) = - \frac{(\alpha_{R;v_i} - \beta_{D;v_i}) |e|}{\hbar \mu_{\rm B}} \langle z\rangle
\label{g-factor-change}
\end{equation}
[for a discussion of the gauge-invariance of this result, see
Appendix \ref{App C: Gauge-invariance}].
The $g$-factor scales as $F_z^{2/3}$, which is close to a linear scaling over the range ($\sim 6\%$) of
the experimentally applied electric fields, see Fig.~\ref{fig:3}b.

Since the in-plane $g$-factor correction, $\delta g_{\parallel}^{v_i}$, is proportional
to $\alpha_{R;v_i}$, $\beta_{D;v_i}$ it is clear that 
for the two eigenvalley subspaces it may change sign
along with the sign change of
$\alpha_{R;v_i}$, $\beta_{D;v_i}$, Eq.~(\ref{sign-SOC}).
E.g.,
for the intra-valley spin-flip parameters being exactly zero,
$s_R, s_D = 0$, the $g$-factor
correction
will be exactly opposite
\begin{equation}
\delta g_{\parallel}^{v_1} = - \delta g_{\parallel}^{v_2}
\label{opposite_change} .
\end{equation}
Relatively smaller corrections due to non-zero intra-valley spin flipping,
$s_R, s_D \neq 0$ will generally violate Eq.~(\ref{opposite_change}),
leaving the $g$-factor
corrections
opposite in sign, but with different absolute value,
$|\delta g_{\parallel}^{v_1}| \neq |\delta g_{\parallel}^{v_2}|$, which is observed in the current experiment\cite{VeldhorstRuskov2015PRB},
see Fig.~\ref{fig:3}.
Tight-binding calculations\cite{Nestoklon3-2008} were performed for the case of a Si/SiGe interface,
with the result that
$|\chi_D| \gg |s_D|$, $|\chi_R| \gg |s_R|$, while $|\chi_R| \sim |s_D|$,
supporting the case of Eqs.~(\ref{sign-SOC}) and (\ref{opposite_change}).
For comparison of the results Eqs.~(\ref{g-eff-approx})-(\ref{g-factor-change})
with the experiment, see Sec. \ref{sec: comparison_with_experiment}.

\subsubsection{$\delta g_{\parallel}$ to 2nd-order PT}
\label{sec: 2nd-order-g-parallel}

Since at certain angles of the in-plane magnetic field, Eq.~(\ref{optimal-angle-approx}),
the g-factor 1st-order correction can be zeroed,
one needs to calculate also
higher order effects, which arise due to QD's energy level structure.

We consider a small quantum dot (QD)  in  MOS   $\rm Si/SiO_2$  heterostructure,
Figs.~\ref{fig:1}c and d.
Thus, the QD is designed such that the first excited orbital state for one-electron QD is
at $\Delta_{\rm orb} \simeq 8 {\rm meV}$ above the ground state,
and for the three-electron QD, $\Delta_{\rm orb} \simeq 2 {\rm meV}$ \cite{Yang2013NC}.
Since the valley splitting, $E_{\rm VS}$, between the lowest valley
eigenstates $|v_1\rangle$ and $|v_2\rangle$
is of the order of few hundred $\mu{\rm eV}$ in such heterostructures
the structure of levels  is that shown on Figs.~\ref{fig:1}c and d,
with the two closely spaced eigenvalley states,
separated by $\Delta_{\rm orb} \equiv \hbar\omega_0 \gg E_{\rm VS}$  from the first two orbital excited QD states
(Appendix \ref{App B: QD level structure corrections}).
The shorthand notation
$|\bar{v}_i\rangle \equiv |v_i\rangle \otimes |\phi_{v_i}(\bm{r})\rangle$, $i=1,2$, includes
the eigenvalley state and the envelope wavefunction
$|\phi_{v_i}(\bm{r})\rangle \equiv \phi_0^{v_i}(x,y) \varphi_0(z)$ 
of the electron confined in the QD.
The envelope wave function  may depend on the valley index
for a non-ideal interface (with roughness) \cite{Yang2013NC,Gamble2013PRB}.
Similarly, the states
$|m_1\rangle \equiv |v_1\rangle \otimes |0_x,1_y,0_z\rangle$ and
$|m_2\rangle \equiv |v_1\rangle \otimes |1_x,0_y,0_z\rangle$, and
$|\tilde{m}_1\rangle \equiv |v_2\rangle \otimes |0_x,1_y,0_z\rangle$ and
$|\tilde{m}_2\rangle \equiv |v_2\rangle \otimes |1_x,0_y,0_z\rangle$, 
include first orbitally excited states.
The states $|m_1\rangle$, $|m_2\rangle$
as well as $|\tilde{m}_1\rangle$, $|\tilde{m}_2\rangle$
are degenerate for a circular QD \cite{FriesenCoppersmith2010},
and split from each other by $E_{\rm VS}$.
We will neglect higher orbital excitations, assuming parabolic lateral confinement, Fig.~\ref{fig:1}d.

In a magnetic field each of these levels are Zeeman split, with $E_Z  = g^* \mu_B B$,
and we enumerate them as $|1\rangle,|2\rangle\dots ,|12\rangle$
(e.g., $|1\rangle \equiv |\bar{v}_1,\downarrow\rangle$, $|2\rangle \equiv |\bar{v}_1,\uparrow\rangle$,
$|3\rangle \equiv |\bar{v}_2,\downarrow\rangle$, $|4\rangle \equiv |\bar{v}_2,\uparrow\rangle$,
$|5\rangle \equiv |m_1,\downarrow\rangle$, $|6\rangle \equiv |m_1,\uparrow\rangle$, etc.).
%
In fact, $|2\rangle =|\bar{v}_1,\uparrow \rangle$  and  $|3\rangle =|\bar{v}_2,\downarrow \rangle$
anti-cross at $E_Z  = E_{\rm VS}$  (for notations see below and in Appendix \ref{App B: QD level structure corrections})
with energy splitting \cite{Yang2013NC,Xiao2014NC}
$2|V_{23}| \equiv \Delta_a \simeq \frac{\sqrt{2} m_t E_{\rm VS} |\beta_{D;21} - \alpha_{R;21}|}{\hbar} (x_{12} + y_{12})$
in the presence of interface roughness \cite{Yang2013NC,Xiao2014NC},
and due to the effective  Rashba and Dresselhaus
SOC interaction Hamiltonians, ${\cal H}_{\rm s-v}^{ij}$
Eq.~(\ref{spin-valley-mixing-3D}).
%
%
Using this level structure,
one is able to describe successfully the experimentally observed ``relaxation hot spot'' that
occurs in the region of maximal spin-valley mixing \cite{Yang2013NC}, at $E_Z \approx E_{\rm VS}$
(where the phonon relaxation is strong).
Moreover, the standard SOC corrections via the virtual excitation to
the orbital levels $|m_{1,2}\rangle$  describe correctly the $B^7$ magnetic field dependence
of the relaxation rate above the anticrossing \cite{Yang2013NC}, at $E_Z > E_{\rm VS}$.
(For a three-electron QD, the structure of levels is essentially the same, Fig. 1c:
this explains essentially the experimentally identical ``relaxation hot spot''
measured in the 3e-system \cite{Yang2013NC}).

For the 2nd-order correction to the g-factor of the lower valley ($v_1$) electron,
$\delta^{(2)} g^{v_1}_{\parallel} = [\delta E_2^{(2)} - \delta E_1^{(2)}] /(\mu_B B_{\parallel})$,
we use standard perturbation theory for the energy
difference  $[\delta E_2^{(2)} - \delta E_1^{(2)}]$
(Appendix \ref{App B: B-parallel}).
\begin{eqnarray}
&& \delta E_2^{(2)} - \delta E_1^{(2)} = \frac{2|V_{12}|^2}{E_Z} +
|V_{14}|^2 \left( \frac{1}{E_Z - E_{\rm VS}} + \frac{1}{E_Z + E_{\rm VS}}\right)
\nonumber\\
&& \quad\quad {} + 2|V_{16}|^2 \left( \frac{1}{E_Z - \Delta_{\rm orb}} + \frac{1}{E_Z + \Delta_{\rm orb}}\right)
\nonumber\\
&& \quad {} + 2|V_{1,10}|^2 \left( \frac{1}{E_Z - \Delta_{\rm orb}-E_{\rm VS}} + \frac{1}{E_Z +
\Delta_{\rm orb}+E_{\rm VS}}\right)
\label{delta-g-parallel-energy-main}  .
\end{eqnarray}
The matrix elements $V_{ab},\ a=1,2,\,b=1,\ldots,12$, are routinely
calculated,
using the relation between
matrix elements of momentum and position
via the equation of motion.
%
In Eq.(\ref{delta-g-parallel-energy-main}) we have used that $V_{23}=V_{14}$, $V_{25}=V_{16}$,
$V_{27}=V_{18}$, etc.,
and also that $V_{16}=V_{18}$, $V_{1,10}=V_{1,12}$ for a circular dot
(Appendix \ref{App B: B-parallel}).
SOCs, Eq.~(\ref{spin-valley-mixing-3D}),  make the qubit states,
$|1\rangle \equiv |\bar{v}_1,\downarrow \rangle$,
$|2\rangle \equiv|\bar{v}_1,\uparrow \rangle$, to mix with the
upper orbital states $|m_1\rangle$, $|m_2\rangle$,
$|\tilde{m}_1\rangle$, $|\tilde{m}_2\rangle$,
as well as with the $|\bar{v}_2\rangle$-states.
The mixing to the $|\bar{v}_2\rangle$-states (which have a quasi s-like envelope)
is via the transition dipole
matrix elements
$\bm{r}_{12} \equiv \langle \bar{v}_1| \bm{r} |\bar{v}_2\rangle$
(notice, $\bm{r}_{12} \neq 0$ only due to roughness effects \cite{Yang2013NC,Gamble2013PRB}),
and the mixing to the higher orbital states
$|m_i\rangle$, $|\tilde{m}_i\rangle$ is via the
standard orbital dipole
matrix elements,
i.e.,
$\bm{r}_{1,m_1} \equiv  \langle v_1| \bm{r} |m_1\rangle$, etc.;
for a circular dot:
$y_{1,m_1}=x_{1,m_2} = \sqrt{\frac{\hbar^2}{2 m_t \Delta_{\rm orb}}}$
(also, we assume $y_{1,\tilde{m}_1}=x_{1,\tilde{m}_2}\simeq y_{1,m_1}$).

Here we present the approximate result (for exact expressions, see Appendix \ref{App B: B-parallel}),
assuming  $x_{12}=y_{12} \sim \langle z\rangle \simeq {\rm few\ nm}$,
and SOC constant relations suggested by the tight-binding calculations:
$\alpha_{R;v_1} \ll \beta_{D;v_1}$, and $|\alpha_{R;21}| \ll |\beta_{D;21}|$.
%
For the relevant (to the experiment) case of $E_Z \ll E_{\rm VS}, \Delta$
one gets
\begin{eqnarray}
&&  \delta^{(2)} g^{v_1}_{\parallel}  \simeq  \frac{|e|^2}{\hbar^2 \mu_B^2} \left\{
\beta_{D;v_1}^2 \cos^2 2\varphi \, \langle z \rangle^2
-\left( m_t/m_0 \right)^2 \times
\right.  \qquad\qquad\quad
\nonumber\\
&&
\quad
\left.  \times \left[ |\beta_{D;21}|^2 (1+\sin 2\varphi) \, x_{12}^2 +
        \left(\beta_{D;v_1}^2 +  |\beta_{D;21}|^2 \right) \, y_{1 m_1}^2
        \right]   \right\}.
\label{dg2-parallel}
\end{eqnarray}
In Eq.~(\ref{dg2-parallel}) the first term ($\sim \langle z\rangle^2$)
is exact and can be extracted from the second order expansion of Eq.(\ref{g_eff})
for $v_i = v_1$
[it is zero in the $[110]$ direction].
It can be seen that the whole 
2nd order correction
is of the order of
$|\delta^{(2)} g^{v_1}_{\parallel}| \sim [\delta^{(1)} g^{v_1}_{\parallel} ]^2 \sim 10^{-6}$.
(We assume that similar relation holds for the $v_2$-electrons, without calculation).

The smallness of the second order contribution can be also seen
by noting that
the second term ($\sim x_{12}^2$) and the third term ($\sim y_{1 m_1}^2$)
in Eq.~(\ref{dg2-parallel}) are  proportional to the small ratios
$|\Delta_a|^2/E_{\rm VS}^2$ and $|V_{16}|^2/\Delta_{\rm orb}^2 = m_t |\beta_{D;21} - \alpha_{R;21}|^2/(4 \Delta_{orb})$
that are of the order of $10^{-6}-10^{-8}$,
since the splitting at
the spin-valley
anticrossing is small \cite{Yang2013NC,Xiao2014NC}, $\Delta_a \approx (10^{-3}-10^{-4}) E_{\rm VS}$.

At the spin-valley
anticrossing, $E_Z \approx E_{\rm VS}$, the $g$-factor change is somewhat bigger,
$|\delta g| \sim \Delta_a / E_{\rm VS}$,
which is still at least one order of magnitude smaller than is experimentally observed.
Moreover, the electric field dependence in $F_z$ arising from this contribution is non-linear,
which is not observed experimentally \cite{VeldhorstRuskov2015PRB}
(Appendix \ref{App B:  delta g at anticrossing}).
This experimental fact restricts the size of the spin-valley splitting at the anticrossing point \cite{Yang2013NC}.
Also notice that due to quadratic dependence on the SOC constants
this contribution
would be insensitive to the change of
their sign.

\subsection{g-factor for perpendicular magnetic field, $\bm{B}_{\perp}$}

\subsubsection{$\delta g_{\perp}$ to 1st-order PT}

For a perpendicular magnetic field one chooses  the gauge $\bm{A}_{\perp}(\bm{r}) = \frac{B_z}{2}(-y, x, 0)$;
In what follows,
we again neglect small corrections originating from the bulk Hamiltonian ${\cal H}_0$, Eq.~(\ref{bulk_H}).
The perturbation to Eq.~(\ref{H-vi}), $\delta_B {\cal H}_{v_i}^{\rm 3D}$,
due to perpendicular magnetic field $\bm{B}_{\perp}$, contributes to $\delta g_{\perp}$
to first order.
%
Averaging it  over the states $|\bar{v}_i\rangle$
as in Eq.~(\ref{H_B_parallel})
gives
\begin{eqnarray}
&&
\langle \bar{v}_i | \delta_B {\cal H}_{v_i}^{\rm 3D} |\bar{v}_i\rangle =
\frac{|e|}{\hbar \mu_B} \ \mu_B \frac{B_z}{2}
\Big[  \alpha_{R;v_i} (\sigma_x  x_{11} + \sigma_y  y_{11}) \Big.
\nonumber\\
&& \qquad\qquad\qquad \Big. { } - \beta_{D;v_i} (\sigma_x  y_{11} + \sigma_y  x_{11}) \Big],\ \ i=1,2
\label{H_B_perp} .
\end{eqnarray}
Similar to Eq.(\ref{total_Zeeman})
the total Zeeman energy can be written via the g-factor tensor:
\begin{equation}
{\cal H}^{\rm tot}_{\rm Z,v_i} =
\mu_B \frac{B_z}{2} \left( g_0 \sigma_z + \delta g^{v_i}_{xz} \sigma_x +
\delta g^{v_i}_{yz} \sigma_y  \right)
\label{total-Zeeman-perp} ,
\end{equation}
where
\begin{eqnarray}
&& \delta g^{v_i}_{xz} = \frac{|e|}{\hbar \mu_B} \left( \alpha_{R;v_i} x_{11} - \beta_{D;v_i} y_{11} \right)
\label{delta-g-xz}\\
&& \delta g^{v_i}_{yz} = \frac{|e|}{\hbar \mu_B} \left( \alpha_{R;v_i} y_{11} - \beta_{D;v_i} x_{11} \right)
\label{delta-g-yz} ,
\end{eqnarray}
and $\bm{r}_{11} \equiv \langle \bar{v}_1| \bm{r} |\bar{v}_1 \rangle$.
These contributions would be zero for an ideal interface, while they
may be non-zero for an interface with roughness, e.g.,
due to
atomic steps\cite{Yang2013NC,Gamble2013PRB}.
In fact, just these matrix elements are needed in order to explain the
``relaxation cold spot'' for a QD with two electrons \cite{Yang2013NC}.
The first-order correction, however, is zeroed as the perturbation is off-diagonal in spin.

\subsubsection{$\delta g_{\perp}$ to 2nd-order PT}
\label{sec: 2nd-order-g-perp}

Exact diagonalization of (\ref{total-Zeeman-perp}) allows to extract
a partial second order contribution, similar to Eqs.~(\ref{g_eff}) and (\ref{dg2-parallel}):
\begin{eqnarray}
&& \delta g_{\perp}^{v_i} =
\frac{|e|^2}{\hbar^2 \mu_B^2} \frac{1}{2g^*}
\left\{
\left( x_{11}^2 + y_{11}^2 \right)\, \left( \alpha_{R;v_i}^2 + \beta_{D;v_i}^2 \right) \right.
\nonumber\\
&& \left. \qquad\qquad\qquad
{ } - 4 x_{11} y_{11} \alpha_{R;v_i} \beta_{D;v_i}
\right\} .
\label{dg2-perp-1st}
\end{eqnarray}
Adding the contributions of the higher levels and
using the same approximations as in
subsection \ref{sec: 2nd-order-g-parallel}, just before Eq.~(\ref{dg2-parallel}),
we obtain (Appendix \ref{App B: B-perpendicular}):
\begin{eqnarray}
&&  \delta^{(2)} g^{v_1}_{\perp} \simeq
\frac{|e|^2}{\hbar^2 \mu_B^2}
\left\{
\beta_{D;v_1}^2 \frac{x_{11}^2}{2}
- 2 \left( m_t/m_0 \right)^2 \left( m_0/m_t -1 \right)\times
\right.
\nonumber\\
&&
\qquad\qquad
\left.  \times \left[ \beta_{D;v_1}^2 \,
                 \left( x_{12}^2 + y_{1 m_1}^2 \right)
               + |\beta_{D;21}|^2 \, y_{1 m_1}^2
                \right]   \right\}.
\label{dg2-perp}
\end{eqnarray}
In Eq.~(\ref{dg2-perp}) the first term ($\sim x_{11}^2$)
is exact and is taken from Eq.~(\ref{dg2-perp-1st}).
It can be seen again that the whole expression is of the order of
$|\delta^{(2)} g_{\perp}| \sim [\delta^{(1)} g_{\parallel} ]^2 \sim 10^{-6}$.

\subsection{g-factor total angular dependence}
To leading order in $a[U_z]$, and neglecting the contributions,
Eqs.~(\ref{delta-g-xz})  and (\ref{delta-g-yz}),
the effective $g$-factor correction is obtained from
Eqs.~(\ref{H_B_parallel})  and (\ref{total-Zeeman-perp})
and reads:
\begin{equation}
\delta g^{v_i}(\varphi, \theta) \simeq
- \frac{|e|}{\hbar \mu_B} \frac{\langle z \partial_z U_z \rangle}{\langle \partial_z U_z \rangle}\,
\left( \alpha_{R;v_i} - \beta_{D;v_i} \sin 2\varphi \right) \sin^2 \theta
\label{g-factor-leading-angular-dependence} ,
\end{equation}
where the magnetic  field components are chosen as: $\bm{B} = B (\sin\theta \cos \varphi, \sin\theta \sin \varphi, \cos \theta)$.
Corrections from the  matrix elements,
Eqs.~(\ref{delta-g-xz})  and (\ref{delta-g-yz}),
give an additional contribution with a different angular dependence:
\begin{equation}
\delta g^{v_i}_{\perp}(\varphi, \theta) \simeq
- \frac{1}{2} \sin 2 \theta \left( \cos \varphi \,\, \delta g_{xz}^{v_i} + \sin \varphi \,\, \delta g_{yz}^{v_i} \right)
\label{g-factor-correction-to-angular-dependence} .
\end{equation}
However, the preservation of the $C_{\rm 2v}$-symmetry would exclude roughness/steps within the dot,
thus eliminating the latter contribution.

\subsection{Discussion of the results and comparison to experiment}
\label{sec: comparison_with_experiment}
%
\subsubsection{Angular dependence}
Our predicted $g$-factor angular dependence (see Fig.~\ref{fig:2}) of the leading contributions
for an applied magnetic  field,  both in-plane, Eq.~(\ref{g-eff-approx}),
and perpendicular to the interface,
Eq.~(\ref{g-factor-leading-angular-dependence}),
was recently
confirmed in an experiment using Si-MOS DQD structure \cite{Sandia-experiment1}.
In the DQD experiment \cite{Sandia-experiment1} the Singlet-Triplet qubit is manipulated
via the  energy detuning between the dots which translates in different perpendicularly applied electric fields
at each dot, and therefore to a different $g$-factor, Eq.~(\ref{g-eff-approx}).
The measured angular dependence,
both in-plane and out-of-plane,
is compatible with the predicted $\sim \sin{2\varphi} \sin^2 \theta$
angular dependence of Eq.~(\ref{g-factor-leading-angular-dependence})
[see also Eq.~(\ref{g-eff-approx})].
The
angle $\varphi_{v_1}$, Eq.~(\ref{optimal-angle-approx}), at which the $g$-factor   
correction is zero,
allows essentially to extract the ratio of the Dresselhaus vs. Rashba constants   
for the lowest eigenvalley band $v_1$: $\beta_{D;v_1}/\alpha_{R;v_1} \approx 8.3$,
at the conditions of the experiment \cite{Sandia-experiment1}.
The smallness of  the calculated by us  second-order corrections to the $g$-factor,
Eqs.~(\ref{dg2-parallel})  and (\ref{dg2-perp}),
including that coming from the QD level structure,
is consistent both with the single QD experiment \cite{VeldhorstRuskov2015PRB}
and with the recent DQD experiment \cite{Sandia-experiment1,TanttuDzurak-experiment}.

\subsubsection{Valley dependence}
While the single QD experiment \cite{VeldhorstRuskov2015PRB} was performed for a fixed in-plane magnetic field
along the crystallographic $[110]$-direction, it has revealed important information about the
valley dependence of the $g$-factor, predicted in
Eqs.~(\ref{g-eff-approx}) and (\ref{g-factor-change}).
%
Indeed, because of the strong lateral confinement
the orbital splitting is much larger than the valley splitting:
$\Delta_{\rm orb} \gg  E_{\rm VS}$,
and it is now clear that if the Si QD is occupied by a single electron,
then one is measuring
the $g_{\rm v_1}$-factor of the lower valley state, $|v_1\rangle$, Fig.~\ref{fig:1}c, left.
For a QD occupied by 3 electrons, Fig.~\ref{fig:1}c, right, the ``valence'' electron is at the
upper valley eigenstate $|v_2\rangle$, and thus $g_{\rm v_2}$ is effectively measured.
Despite the smallness of the $g$-factor change
as a function of the applied electric  field,
the corresponding energy change can be resolved since it happens to be
$\sim 3000$ times larger than the corresponding ESR line width of $2.4\,{\rm kHz}$.
%
The electric field dependence allows the spin qubit evolution to be switched on/off by tunning it in/out of resonance
with an external microwave drive\cite{Veldhorst2014Nn,VeldhorstRuskov2015PRB}.

\begin{figure} [t!]
    \centering
        \includegraphics[width=0.5\textwidth]{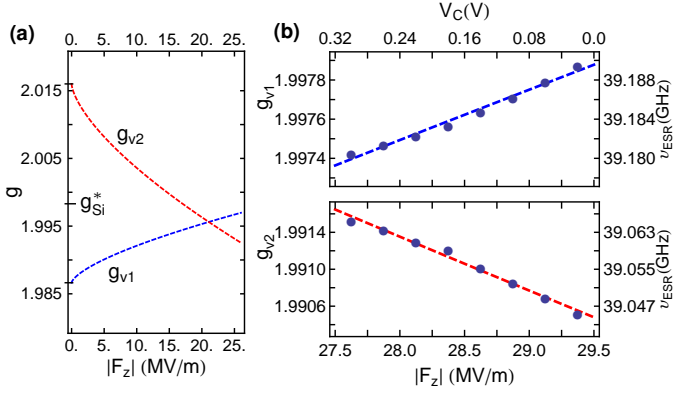}
        \caption{In (a) the valley $g$-factors are plotted depending on the
        electric field $F_z$, based on the model,
        Eqs.~(\ref{g-eff-approx}), (\ref{g-factor-change}),  and (\ref{g-factor-Fz-dependence}),
        for each eigenvalley state:
        $|v_1\rangle$, $|v_2\rangle$.
        The $g$-factor offsets at zero  field and the intercept at $F_z^{\rm int} \simeq 2.1\times 10^7\, {\rm V/m}$ are shown.
        In (b) the experimental electrical control over the valley $g$-factors is shown
        (dots, at a fixed magnetic field $B_0 = 1.4015\, {\rm T}$).
        The experimental points are fitted by the model, Eq.~(\ref{g-factor-Fz-dependence}).
        Tuning both the confinement gate and the plunger gate at the QD \cite{VeldhorstRuskov2015PRB}
        provides control of the
        electric field $F_z$,
        and with that we can vary the qubit resonance frequency
        over several MHz.
        The experimentally observed opposite dependence of the valley $g$-factors on the electric field
        is attributed to the mixing of the original bulk degenerate spin-valley wave-functions
        at the $\rm Si/SiO_2$ interface, via the dominance of the inter-valley spin-flip contributions in the BC,
        Eqs.~(\ref{BC})  and (\ref{spin-valley-mixing}).
        }
        \label{fig:3}
\end{figure}

Let us perform a rough estimation of the 2D spin-orbit parameters, $\alpha_{R;v_i}$, $\beta_{D;v_i}$,
based on the measured $g$-factor dependencies, Fig.~\ref{fig:3}b
and using the predicted
electric field dependence
in the range of high electric fields,
$\delta g^{\rm th}_{\rm v_i} = A_{v_i} F_z^{2/3}$,
Eqs.~(\ref{g-factor-scaling})  and (\ref{g-factor-change}).
The measured change of the $g(F_z)$-factors is approximately a linear function of the electric field $F_z$
for the experimental electric field range, $F_z \approx (2.75 - 2.95)\times 10^7\, {\rm V/m}$,
and
$g_{\rm v_1}(F_z)$ grows with the increasing of $F_z$ (Fig.~\ref{fig:3}b, upper panel),
while $g_{\rm v_2}(F_z)$  decreases (Fig.~\ref{fig:3}b, lower panel).
The experimental energy change of $10 - 20\, {\rm MHz}$ corresponds to
a  $g$-factor changes,
$\Delta [\delta g_{\rm v_1}], \Delta [\delta g_{\rm v_2}] \approx 10^{-3}$.
Moreover, the  measured $g$-factor changes {\em are opposite in sign}, and fulfill the approximate relation
\begin{equation}
\Delta [\delta g^{\rm exp}_{\rm v_2}(F_z)] \simeq - 2.24\  \Delta [\delta g^{\rm exp}_{\rm v_1}(F_z)]
\label{Delta-g-v1-v2-exp} ,
\end{equation}
which was qualitatively explained in Sec. \ref{sec: g-parallel}
via the {\em dominance of  the  inter-valley spin-flip scattering amplitudes}
over the intra-valley spin-flip amplitudes
in the BC, Eq.~(\ref{BC}).
Since
$\Delta [\delta g^{\rm exp}_{\rm v_2}(F_z)] / \Delta [\delta g^{\rm exp}_{\rm v_1}(F_z)]  =
A_{v_2} /A_{v_1}$ (for high fields),
one can extract the ratio:
\begin{equation}
\frac{A_{v_2}}{A_{v_1}} =
\left.
\frac{\alpha_{R;v_2} - \beta_{D;v_2}}{\alpha_{R;v_1} - \beta_{D;v_1}}
\right|_{\rm high-field, \varphi = \frac{\pi}{4}}
\simeq - 2.24
\label{ratio-V2-to-v1} .
\end{equation}
Moreover, expanding $\delta g_{\rm v_i}$ to second order:
$\Delta [\delta g_{\rm v_i}] \simeq  A_{v_i} F_z^{2/3} \frac{2}{3} (\frac{\Delta F_z}{F_z} - \frac{1}{3} \frac{\Delta F_z^2}{F_z^2})$,
with $\Delta F_z = 0.175\times 10^7 \,{\rm V/m}$ (Fig.~\ref{fig:3}b),
and using
Eqs.~(\ref{g-eff-approx})  and (\ref{g-factor-change})
one obtains
\begin{eqnarray}
&&\alpha_{R;v_1} - \beta_{D;v_1} \simeq -361\times 10^{-13}\, {\rm eV\cdot cm},    
\label{v1-alpha-beta}
\\
&&\alpha_{R;v_2} - \beta_{D;v_2} \simeq 810\times 10^{-13}\, {\rm eV\cdot cm} ,   
\label{v2-alpha-beta}
\end{eqnarray}
(with a relative error of $5\times 10^{-4}$; however a systematic error due to
deviation from the high-field behavior, $\sim F_z^{2/3}$, is not accounted).
These values are compatible with
qualitative estimations for GaAs  heterojunctions \cite{RashbaSheka-book1991},
and also with
tight-binding calculations of Nestoklon {\em et al.}\cite{Nestoklon3-2008} for a Si/Ge interface.
They are larger than the latter by a factor of 10,
which is expected since here the electric field is $\sim 3$ times higher
than in that calculations,
and the $\rm Si/SiO_2$ interface is more abrupt.

Finally, we would like to stress that the $g$-factor dependence of $F_z^{2/3}$ is for a high
electric field
(see Sec. \ref{sec: valley-splitting}).
Thus, we will model the low-field dependence in a simplistic way,
by adding a (valley dependent) $g$-factor offset $\delta x_{v_i}$ (Fig.~\ref{fig:3}a):
\begin{equation}
g_{\rm v_i} = g_{\rm Si}^* + \delta x_{v_i} + A_{v_i} F_z^{2/3}
\label{g-factor-Fz-dependence} ,
\end{equation}
where $g_{\rm Si}^* \simeq 1.9983$ is the bulk value in Si for in-plane magnetic field
\cite{WilsonFeher1960BullAPS,Roth1960PR,Liu1962PR}.
By fitting Eq.~(\ref{g-factor-Fz-dependence}) to the experimental data, Fig.~\ref{fig:3}b,
one obtains the $g$-factor offsets
%
$\delta x_{v_1} \simeq -0.012$,
and
$\delta x_{v_2} \simeq 0.018$
(with an  
error of $5\times 10^{-4}$),
for this particular angle $\varphi = \pi/4$, when $B_{\|}$  is along the $[110]$-direction.
We note, that  the assumed  $C_{\rm 2v}$-symmetry of the interface (quantum well)
implies that the low-electric field Hamiltonian will be described by the same
invariant Rashba and Dresselhaus structures, see Eq.~(\ref{spin-valley-mixing-3D}).
This would imply some $\sim \sin 2\varphi$ dependence of the offset values, reflecting the symmetry.
A theory of the low-electric field effects in the $g$-factors, including  offsets
will be considered elsewhere.

While an interface with roughness (which is a realistic interface)
will generally violate the ``global'' $C_{\rm 2v}$-symmetry,
one might expect, for relatively small dots,
a situation when the $C_{\rm 2v}$-symmetry is not violated within the quantum dot.
This symmetry will dictate the form of the interface Hamiltonian, e.g in Eq.~(\ref{spin-valley-mixing-3D}),
and the $g$-factor angular dependence,
derived in
Eqs.~(\ref{g-eff-approx})  and (\ref{g-factor-leading-angular-dependence}).
This physical intuition was recently confirmed experimentally, by observing the
angular dependence in a Si-MOS DQD experiment \cite{Sandia-experiment1,TanttuDzurak-experiment}.
Similar angular dependence was also revealed in a single QD with micromagnet,
manipulated  at a Si/Ge interface \cite{Delft-experiment}.
We stress that any explicit violation of the $C_{\rm 2v}$-symmetry,
(e.g., via explicit atomic step in the QD \cite{UNSW-experiment-steps})
will not result in the angular dependence predicted here for the g-factor, Eq.~(\ref{g-eff-approx});
moreover, one would not be allowed to speak about Rashba and Dresselhaus contributions in
the Hamiltonian. More experimental and theoretical work is needed to understand the role
of atomic steps/roughness on the g-factor and other parameters.

\subsubsection{Spin-orbit coupled electric field noise}
\label{sec: new_noise_mechanism}

The $F_z$-dependence of the $g$-factor implies that a new dephasing mechanism
is introduced via the fluctuations of the (gate) electric field,
which was discussed in the context of 1e- and 3e- qubit
using randomized benchmarking sequences to reveal it \cite{Magesan2012PRL,Fogarty2015PRA,VeldhorstRuskov2015PRB}.
For the single QD qubit of Ref.~\onlinecite{VeldhorstRuskov2015PRB}  this is the
detuning noise
$\delta \epsilon(t)$ of the Hamiltonian
${\cal H}_{Qb} = \frac{\epsilon}{2} \sigma_z + \frac{B_1^{\rm ac}}{2} \sigma_x$,
where $\epsilon \propto \nu_{\rm ESR} - \nu_{v_i}$ is the detuning,
and $B_1^{\rm ac}$ is the ac driving amplitude.
Assuming a white noise,
$\delta\epsilon(t) = \xi_{\epsilon}(t)$
with a (single-sided) noise spectral density $S_{\epsilon}$,
(see, e.g. Ref.~\onlinecite{Korotkov-nonideal2003PRB}),
the dephasing rate $\gamma_{v_i}$ is derived
at a chosen field $F_{z0}^{*}$ as
%
\begin{equation}
%
\gamma_{v_i} = \frac{S_{\epsilon}}{4 \hbar^2} =
\frac{(\mu_B B_{\parallel})^2}{4 \hbar^2} \left|\frac{\partial[\delta g_{\parallel}^{v_i}(\varphi,F_{z0}^{*})]}{\partial F_z}\right|^2 \, a^2\, S_V ,
\label{new_charge_noise}
\end{equation}
where $S_{\epsilon}$
is linearly related
to the gate voltage spectral density $S_V$,
assuming linear dependence of field vs. voltage, $F_z \equiv  a\, V$ (see, Fig.~3b and Ref.~\onlinecite{Yang2013NC}).
%
From Eq.~(\ref{g-factor-derivative}) one  obtains suppression for high fields,
e.g., for a linear confinement:
$\frac{\partial[\delta g_{\parallel}^{v_i}(\varphi,F_z)]}{\partial F_z} = \frac{1}{F_z}\, \delta g_{\parallel}^{v_i}(\varphi,F_z)$.
Using Eq.~(\ref{Delta-g-v1-v2-exp}), the dephasing rates for the 3e and 1e qubits
(for $\varphi = \pi/4$)
should be related as:
$\gamma_{v_2}  \simeq (2.24)^2\,  \gamma_{v_1}$.
On the other hand, using Hahn echo measurements one can cancel out the $1/f$ (drift) noise,
and
the measured $T_2$ reveals:
$T_2^{3e} \approx 400 \mu{s}$ and  $T_2^{1e} \approx 1200 \mu{s}$, i.e.
a dephasing rate ratio of $3$
instead of $(2.24)^2$.
This can be explained assuming another (valley-independent) dephasing $\gamma_0$
(it can be associated with some charge fluctuators or noise on the ac amplitude $B_1^{\rm ac}$).
Thus,
$\gamma_{3e} = \gamma_{v_2} + \gamma_0$,
$\gamma_{1e} = \gamma_{v_1} + \gamma_0$, with $\gamma_0 \simeq \gamma_{v_1}$,
i.e. $\gamma_0$ is comparable to $\gamma_{v_1}$ in this experiment.

The quadratic dependence of the noise   
on the $g$-factor change:
$S_{\epsilon} \propto [\delta g_{\parallel}^{v_i}(\varphi,F_z)]^2$, Eqs.~(\ref{g-factor-derivative}) and (\ref{new_charge_noise}), implies that
it can be zeroed at the ``sweet spot angles'' $\varphi_{v_i}$, defined in
Eq.~(\ref{optimal-angle-approx}).
At these angles (which may be different for the two eigenvalley subspaces, $v_1, v_2$)
either $\gamma_{1e}$ or $\gamma_{3e}$ will take the minimal value $\gamma_0$.
Similar decrease of the noise can be achieved by rotating the field perpendicular to the interface,
since the $g$-factor corrections are strongly suppressed, see
Eqs.~(\ref{dg2-perp-1st})  and (\ref{dg2-perp}).

\section{Summary and Discussion}

This paper presents a detailed theory
to explain measurements of unexpected g-factor shifts in silicon
quantum dots and to predict future experiments and impact to silicon-based quantum computing.
We derived the effective spin-orbit interaction from appropriately formulated boundary conditions
that take into account the symmetry of the
silicon heterostructure interface and the hermiticity of the problem at hand.
These effective spin-orbit interactions are used to derive the valley splitting at the
interface, both its scaling with the applied electric field (perpendicular to the interface)
and with the interface $z$-confinement for the conduction electrons.
Then the 3D (and 2D) effective Rashba and Dresselhaus spin-orbit interactions are calculated,
assuming   
a $C_{2v}$ interface symmetry.
We argue that these  new interface SOC contributions are much
stronger than possible bulk contributions.
Compared to previous phenomenological approaches  
\cite{Tahan2002PRB66,Chutia2007,Goswami2007NP,Shaji2008NP,Saraiva2009PRB,
FriesenCoppersmith2010,Culcer2010PRB,Culcer2010PRB2,RancicBurkard2016PRB,Yang2013NC,Xiao2014NC,Kawakami2014Nn},
the approach taken in this paper
provides more rigorous ground for analyzing current
and future experiments.

The effective spin-orbit interactions contain both diagonal (in the eigenvalley number)
and off-diagonal contributions, which are to be used in the analysis of experiments that involve
both eigenvalley states (e.g., in the so-called valley qubits\cite{Pratti2011JNN,Culcer2012PRL}).
Based on the above, we derived the electron $g$-factors for conduction 2DEG electrons
(at a relatively weak lateral confinement) for an applied in-plane or perpendicular to the
interface magnetic field. To leading order, we predicted the angular dependence of the $g$-factor
with the in-plane angle, as well as with the azimuthal angle (for a magnetic field having a
perpendicular component). For appropriate experiments with a single QD
these predictions would  allow us to extract the ratio of Rashba and Dresselhaus effective constants,
from a measured $g$-factor angular dependence.
In fact, any significant angular dependence will show that the Dresselhaus contribution
dominates the Rashba one, thus supporting our statement that interface contributions
are much stronger than that originating from the bulk.

The physical mechanism that causes shifts in the SOC parameters (and thus g-factor) as a function of electric field
allows a new path for charge noise to affect the qubit.
The predictions in this paper on the $g$-factor angular dependence are made for both
lower and upper eigenvalley subspaces, which in general may have different
spin-orbit (Rashba and Dresselhaus) contributions.
We predict, based on the in-plane angular dependence, the so-called
{\em sweet spots
in the direction of the magnetic field, when the $g$-factor correction, $\delta g$ is zero},
and therefore there is no electric field scaling; consequently, the corresponding spin qubit
would be insensitive (to first order) to the gate voltage (charge noise) of the applied
electric field mediated by these new SOC contributions. As a trivial consequence, a QD
{\em qubit will be also insensitive to gate (charge) noise
when the magnetic field is perpendicular to the interface}, as in this case the $g$-factor
variation
is equally suppressed. To estimate this suppression, we have also calculated the second order corrections
(in the perturbation theory) to the $g$-factor at any magnetic field direction,
which also include the effects of the internal QD level structure, assuming strong confinement
typical for the current experiments \cite{Veldhorst2014Nn,VeldhorstRuskov2015PRB,Xiao2014NC,Sandia-experiment1,TanttuDzurak-experiment}.
We have shown that these corrections are typically small $\sim 10^{-6}$ which supports
the first order results discussed above.
Eventually, an enhancement of these effects is possible near the so-called ``relaxation hot spot''
\cite{Yang2013NC},
where the $g$-factor corrections may reach $\sim 10^{-4} - 10^{-3}$,
however such enhancement was not observed experimentally \cite{Veldhorst2014Nn,VeldhorstRuskov2015PRB}.
The absence of such enhancement may be explained (is consistent) with our theory,
giving further constraints on the interface BC matrix parameters (both of their amplitudes and phases).

The ability to appreciably change the g-factor of an electron via applied voltages
on top-gates offers a new and unplanned-for opportunity for control of silicon quantum dot qubits.
For example, implementing a 2-qubit encoding \cite{Levy2002PRL}
would allow for all-electrical control
without the need for 3-quantum dots, magnetic field or nuclear gradients; this may be relevant
for quantum computing not only in reducing the overhead in qubits but also in gate pulses as,
for example, it has been recently showed that two-qubit encoded gates can be accomplished in
far fewer gates than 2-DFS encodings \cite{AEON2016}.
Further, that one electron and three electron dots
exhibit different behavior (while both still being good qubits), another opportunity exists for
creative quantum dot gate protocols. On the other hand, g-factor tunability can create new mechanisms
for decoherence, especially an increased sensitivity to charge noise. Our theory predicts a means to
remove this channel by magic magnetic field angles (perpendicular for example).
Finally, we note that the above theory should also apply to   
Si/Ge heterostructure quantum dots,
with the caveat that the shift in g-factor will likely be smaller relative to the MOS-interface dots.

{\em Note Added:} Whilst we were preparing our manuscript
\cite{Delft2016-Ruskov-talk,QCPR2015-SanDiego-Tahan}
we became aware of a relevant experiment on a MOS double quantum dot system \cite{Sandia-experiment1}
(and most recently see the experiment \cite{TanttuDzurak-experiment})
at the similar conditions discussed
in our paper, dealing with the lowest eigenvalley states in the DQD.
Namely, their conditions are at an applied perpendicular to the interface electric field
and at a magnetic field applied at various angles (both in-plane and perpendicular).
The new experimental results of Ref.~\onlinecite{Sandia-experiment1} confirm to a large extent our predictions.

Particularly,
(i) the very possibility to manipulate the Singlet-Triplet DQD qubit is via the difference in the
electron $g$-factor in the two dots, which arises in the deep $(1,1)$ regime, where the
electric field applied to each of the dots becomes essentially different (i.e., far from the symmetric/degeneracy point);\\
(ii)  their observed angular dependence, $\sim \sin 2\varphi$
is compatible with our predictions for the lower eigenvalley subspace,
see Eq.~(\ref{g-eff-approx}).\\
(iii)  Since the difference of the  Dresselhaus and Rashba effective spin orbit couplings,
for the two dots, is linear with the dots' electric field difference,
the ratio of $\Delta \beta / \Delta \alpha \approx 8.3$ extracted in the DQD experiment \cite{Sandia-experiment1}
is exactly the ratio of these couplings
(that is independent of the electric field strength)
$\beta_{D;v_1}/\alpha_{R;v_1}$, for the lower eigenvalley subspace, see
Eqs.~(\ref{SOC-consts-a11})  and (\ref{SOC-consts-b11}).\\
(iv) Finally, we mention that the predicted in our paper angular dependence of the dephasing,
having a minimum dephasing rate at the ``sweet spot angles'', Eq.~(\ref{optimal-angle-approx}),
is yet to be measured in a future experiment.

{\bf Acknowledgments:}
A.S.D. acknowledges support from the Australian Research Council (CE11E0001017 and CE170100039)
and the US Army Research Office (W911NF-13-1-0024 and W911NF-17-1-0198).
The views and conclusions contained in this document are those of the authors
and should not be interpreted as representing the official policies, either expressed or implied,
of the Army Research Office or the U.S. Government. The U.S. Government is authorized to reproduce and distribute
reprints for Government purposes notwithstanding any copyright notation herein.



\appendix

\section{Derivation of the effective surface Hamiltonian from boundary conditions}
\label{App A: eff-Hamiltonian}

In this appendix we derive  Eq.(\ref{eff-interface-Hamiltonian}).
Starting with the boundary condition (\ref{BC}) one denotes it as
${\cal B} \Phi\mid_{z=z_0^{+}} = 0$ with ${\cal B} \equiv {\cal B}_1 + {\cal B}_2$, and
${\cal B}_1  \equiv   1 + i R\, k_z$,
${\cal B}_2  \equiv  - R \frac{2m_l}{\hbar^2} V_{\rm if}(\bm{k})$;
$V_{\rm if}(\bm{k})$ being the interface spin-valley mixing matrix.
%
Since $\langle k^2_{x,y}\rangle \ll \langle k^2_z \rangle$ (for a strong z-confinement)
we will consider ${\cal B}_2$ as a perturbation.
In what follows, we will approximately replace the boundary operator ${\cal B}$  by a unitary one
up to higher order corrections:
\begin{equation}
{\cal B} \simeq \Gamma_{BC}
\label{B_Gamma_replace} ,
\end{equation}
with $\Gamma_{BC}\, \Gamma^{\dagger}_{BC} \simeq 1$.

Indeed, to zeroth order we have the BC
${\cal B}_1 \Phi^{(0)}\mid_{z=z_0^{+}} = 0$, see Eq.(\ref{BC_zeroth_order}).
Then it follows that
\begin{equation}
{\cal B}_2\, {\cal B}_1 \Phi^{(0)}\mid_{z=z_0^{+}} = 0
\label{BC_zero_order} ,
\end{equation}
or
\begin{equation}
{\cal B}_2 \Phi^{(0)}\mid_{z=0^{+}} = {\cal B}_2\, (1 - {\cal B}_1) \Phi^{(0)}\mid_{z=z_0^{+}}
\label{BC_zero_order1} .
\end{equation}

Now, to first order one have
\begin{equation}
{\cal B} \Phi\mid_{z=z_0^{+}} = \Big[ {\cal B}_1\, ( \Phi^{(0)} + \Phi^{(1)} ) + {\cal B}_2\, \Phi^{(0)} \Big] \mid_{z=z_0^{+}} = 0
\label{BC_first_order} ,
\end{equation}
or
%
\begin{equation}
\Big[ {\cal B}_1\, \Phi(z) + {\cal B}_2\, (1 - {\cal B}_1) \Phi(z) \Big]\mid_{z=z_0^{+}} = 0
\label{BC_first_order1} ,
\end{equation}
where we have replaced $\Phi^{(0)}$ by $\Phi$ in the second term of Eq.(\ref{BC_first_order})
up to higher order corrections.
The last BC,  Eq.(\ref{BC_first_order1}), can be rewritten in the form
$\Gamma_{BC} \Phi(z)\mid_{z=z_0^{+}} = 0$  where
%
\begin{eqnarray}
&& \Gamma_{BC} = {\cal B}_1 + {\cal B}_2\, (1 - {\cal B}_1) = 1 + i\,\gamma_{BC}
\\
&& \gamma_{BC} \equiv R\, k_z + R^2 \frac{2m_l}{\hbar^2} V_{\rm if}(\bm{k})\, k_z
\label{B_Gamma_unitary} ,
\end{eqnarray}
and  $\Gamma_{BC}$ is an (approximate) unitary operator,
$\Gamma_{BC}\, \Gamma^{\dagger}_{BC} = 1 + {\cal O}(\gamma_{BC}^2)$,
up to higher orders.

Performing now the unitary transform with $\Gamma_{BC}$ as in Eq.(\ref{transformed-H})
the transformed BC is
$\tilde{\Phi}\mid_{z=z_0^{+}} \equiv \Gamma_{\rm BC}\Phi\mid_{z=z_0^{+}} = 0$
and the transformed Hamiltonian reads:
%
\begin{eqnarray}
&& \tilde{{\cal H}} = \Gamma_{\rm BC} {\cal H}_0 \Gamma^{\dagger}_{\rm BC}
\simeq  {\cal H}_0 + \delta {\cal H} + {\cal O}(\gamma_{BC}^2)
\nonumber\\
&& \delta {\cal H} = i \left[\gamma_{BC}, {\cal H}_0 \right]_{-} =
R \partial_z U_z + R^2 \frac{2m_l}{\hbar^2} V_{\rm if}(\bm{k}) \partial_z U_z
\label{transformed1} .
\end{eqnarray}

\section{QD level structure and its contribution to the $g$-factor}
\label{App B: QD level structure corrections}

In order to emphasize the tunneling Hamiltonian representation
implied by
Eq.~(\ref{4-vector}),
we rewrite
the expressions for the lowest eigenvalley states,
Eq.~(\ref{v1-v2-eigenstates}), to the form
%
\begin{eqnarray}
&& |\bar{v}_{i;\sigma}\rangle = \frac{1}{\sqrt{2}}
\left[
\begin{array}{c}
C_\sigma  \\
     0
\end{array} \right]\, \phi^{+z}_{v_i}(\bm{r})
+
\left( \frac{\mp e^{-i\phi_V}}{\sqrt{2}} \right)
\left[
\begin{array}{c}
     0  \\
C_\sigma
\end{array} \right]\, \phi^{-z}_{v_i}(\bm{r}) \quad\ \
\label{v1-v2-states-tunnel-rep}\\
&&  i=1,2 \, ; \ \ \sigma = \uparrow, \downarrow
\nonumber ,
\end{eqnarray}
where the corresponding valley populations are
$\alpha_{+z}^{v_1} = \frac{1}{\sqrt{2}}$, $\alpha_{-z}^{v_1} = - e^{-i \phi_V}\frac{1}{\sqrt{2}}$,
$\alpha_{+z}^{v_2} = \frac{1}{\sqrt{2}}$, $\alpha_{-z}^{v_2} = + e^{-i \phi_V}\frac{1}{\sqrt{2}}$.
Time reversal maintains the relations:
$|\alpha_{+z}^{v_j}| = |\alpha_{-z}^{v_j}|$ and $\phi_{v_j}^{+z}\bm{r}) = \phi_{v_j}^{-z}\bm{r})$. 
For the lowest energy envelopes
$\phi^{+z}_{v_i}(\bm{r}) = \phi^{v_i}_{0}(x,y)\,\varphi_0(z)$
the dependence on the eigenvalley index $v_i$
is due to interface roughness (atomic steps within the dot),
and makes $\phi^{v_i}_{0}(x,y)$ to acquire a $p$-like contribution\cite{Yang2013NC,Gamble2013PRB}.
The corresponding four lowest states
$|v_i\rangle \otimes |0_x,0_y,0_z\rangle \otimes |\sigma\rangle \equiv |\bar{v}_i,\sigma\rangle$,
are enumerated as:
$|1\rangle \equiv |\bar{v}_1,\downarrow\rangle$,
$|2\rangle \equiv |\bar{v}_1,\uparrow\rangle$,
$|3\rangle \equiv |\bar{v}_2,\downarrow\rangle$,
$|4\rangle \equiv |\bar{v}_2,\uparrow\rangle$,
see Sec. \ref{sec: 2nd-order-g-parallel}.
The higher orbital states, Fig.~\ref{fig:1}d,
$|v_i\rangle  \otimes |1_x,0_y,0_z\rangle \otimes |\sigma\rangle $,
$|v_i\rangle  \otimes |0_x,1_y,0_z\rangle \otimes |\sigma\rangle $,
are enumerated
using the notations
$|m_1\rangle \equiv |v_1\rangle \otimes |1_x,0_y,0_z\rangle$,
$|m_2\rangle \equiv |v_1\rangle \otimes |0_x,1_y,0_z\rangle$,
and  $|\tilde{m}_1\rangle$, $|\tilde{m}_2\rangle$ for $v_1 \to v_2$,
namely:
$|5\rangle \equiv |m_1,\downarrow\rangle$,
$|6\rangle \equiv |m_1,\uparrow\rangle$,
$|7\rangle \equiv |m_2,\downarrow\rangle$,
$|8\rangle \equiv |m_2,\uparrow\rangle$,
$|9\rangle \equiv |\tilde{m}_1,\downarrow\rangle$,
$|10\rangle \equiv |\tilde{m}_1,\uparrow\rangle$,
$|11\rangle \equiv |\tilde{m}_2,\downarrow\rangle$,
$|12\rangle \equiv |\tilde{m}_2,\uparrow\rangle$,
see Sec. \ref{sec: 2nd-order-g-parallel}.
The roughness effects for these states are neglected.
Also, higher orbital states are not considered assuming a close-to-parabolic lateral confinement.

We consider the valley diagonal SOC Hamiltonian Eq.~(\ref{H-vi})
in a 3D form
[since the 2D SOC Hamiltonians are generally inconsistent
with the extension of derivatives].
By suitably rotating the axes for an in-plane magnetic field,
$\bm{B}_{\parallel} =(B_x,B_y,0)$
one obtains
\begin{eqnarray}
&& {\cal H}_{v_i} =
\left[ \alpha_{R;v_i}\,
\left\{ \left( s \tilde{\sigma}_x + c \tilde{\sigma}_z\right) P_y
+ \left( c \tilde{\sigma}_x - s \tilde{\sigma}_z\right) P_x
\right\}
\right.
\nonumber\\
&& \left. { } +
\beta_{D;v_i}\,
\left\{ \left( s \tilde{\sigma}_x + c \tilde{\sigma}_z\right) P_y
+ \left( c \tilde{\sigma}_x - s \tilde{\sigma}_z\right) P_x
\right\}
  \right]\, \frac{\partial_z U_z}{\hbar \langle\partial_z U_z \rangle}
\label{H-vi-rotated} ,
\end{eqnarray}
where $s\equiv \sin \varphi$, $c\equiv \cos \varphi$,
$\bm{P} = \hbar \bm{k} + |e| \bm{A}$ are the extended derivatives,
$B_x= B\cos \varphi$, $B_y= B\sin \varphi$,
and  the Pauli matrices along the new axes are
\begin{equation}
\tilde{\sigma}_z = \frac{\sigma_x B_x + \sigma_y B_y}{B}, \ \
\tilde{\sigma}_x = \frac{\sigma_x B_y - \sigma_y B_x}{B}
\label{new-sigma}
\end{equation}
with
$\tilde{\sigma}_z |\uparrow,\downarrow\rangle = \pm |\uparrow,\downarrow\rangle$,
$\tilde{\sigma}_x |\uparrow, \downarrow\rangle =  |\downarrow, \uparrow \rangle$.

Taking the matrix elements $V_{kk} = \langle k| {\cal H}_{v_1} |k \rangle ,\, k=1,2$
one obtains for the first order correction to the g-factor
($U_z = |e| F_z z$ for simplicity):
\begin{equation}
\delta^{(1)}g^{v_1}_{\|} = \frac{V_{22} - V_{11}}{\mu_B B} =
- \frac{|e|}{\hbar \mu_B} \langle z\rangle \left( \alpha_{R;v_1} - \beta_{D;v_1}\, \sin 2\varphi \right)
\label{g-fact-parallel-1}.
\end{equation}

It is straightforward to see that for a 3-electron QD, one can
write the wave function as a Slater determinant
(mean field approximation is implicit \cite{Culcer2010PRB2,Bakker2015PRB}),
where two of the electrons are occupying
the lowest orbital $|\bar{v}_1 \rangle$, and the ``valence'' electron
occupies the upper (split by $E_{\rm VS}$) orbital, $|\bar{v}_2 \rangle$,
Fig.~\ref{fig:1}c.
Then, the matrix element over the 3e wave function
is reduced to a single-particle matrix element of the form:
$V_{kk} = \langle k| {\cal H}_{v_2} |k \rangle ,\, k=3,4$, which leads to the
expression for $\delta^{(1)}g^{v_2}_{\|}$ analogous to Eq.~(\ref{g-fact-parallel-1}),
with the replacement $v_1 \to v_2$.

\subsection{Second order corrections: case of  $\bm{B}_{\parallel}$}
\label{App B: B-parallel}
For the second order corrections it is convenient to introduce
compact notations for the SOC constants,
Eqs.~(\ref{SOC-consts-a11})-(\ref{SOC-consts-b21}):
$a_{ii} \equiv \alpha_{R;v_i}$, $b_{ii} \equiv \beta_{D;v_i},\ i=1,2$, and
$a_{21} \equiv \alpha_{R;21}$, $b_{21} \equiv \beta_{D;21}$.
The second order corrections include transitions to higher states
with different valley content; so, both diagonal and non-diagonal in valley
SOC Hamiltonians, Eq.(\ref{spin-valley-mixing-3D}), contribute:
\begin{equation}
{\cal H}_{\rm s-v}^{ij} =
\frac{a_{ij}}{\hbar}\, \left( \sigma_x P_y - \sigma_y P_x \right) +
\frac{b_{ij}}{\hbar}\, \left( \sigma_x P_x - \sigma_y P_y \right)
\label{spin-valley-B-field}.
\end{equation}
Rotating the axes as above, one obtains for the
first few matrix elements
\begin{eqnarray}
&& V_{12} = \langle \bar{v}_1, \downarrow |{\cal H}_{\rm s-v}|\bar{v}_1, \uparrow \rangle
\nonumber\\
&& { } = \hbar^{-1}\langle \phi^{v_1}(\bm{x})| a_{11} \left( c P_x + s P_y \right) +
b_{11} \left( s P_x + c P_y \right)|\phi^{v_1}(\bm{x}) \rangle \quad
\label{V12}\\
&& V_{13} = \langle \bar{v}_1, \downarrow |{\cal H}_{\rm s-v}|\bar{v}_2, \downarrow \rangle
\nonumber\\
&& { } = \hbar^{-1}\langle \phi^{v_1}(\bm{x})| a_{12} \left( s P_x - c P_y \right) -
b_{12} \left( c P_x - s P_y \right)|\phi^{v_2}(\bm{x}) \rangle \quad
\label{V13}\\
&& V_{14} = \langle \bar{v}_1, \downarrow |{\cal H}_{\rm s-v}|\bar{v}_2, \uparrow \rangle
\nonumber\\
&& { } = \hbar^{-1}\langle \phi^{v_1}(\bm{x})| a_{12} \left( c P_x + s P_y \right) +
b_{12} \left( s P_x + c P_y \right)|\phi^{v_2}(\bm{x}) \rangle \quad
\label{V14} \\
&& V_{15} = \langle \bar{v}_1, \downarrow |{\cal H}_{\rm s-v}|m_1, \downarrow \rangle
\nonumber\\
&& { } = \hbar^{-1}\langle \phi^{v_1}(\bm{x})| a_{11} \left( s P_x - c P_y \right) -
b_{11} \left( c P_x - s P_y \right)|\phi^{m_1}(\bm{x}) \rangle \qquad
\label{V15}\\
&& V_{16} = \langle \bar{v}_1, \downarrow |{\cal H}_{\rm s-v}|m_1, \uparrow \rangle
\nonumber\\
&& { } = \hbar^{-1} \langle \phi^{v_1}(\bm{x})| a_{11} \left( c P_x + s P_y \right) +
b_{11} \left( s P_x + c P_y \right)|\phi^{m_1}(\bm{x}) \rangle\!, \ \ \qquad
\label{V16}
\end{eqnarray}
etc.
The matrix elements $V_{ab},\ a=1,2,\,b=1,\ldots,12$, are routinely
calculated,
using the relation between momentum and position
matrix elements
via the equation of motion.
For example,
\begin{eqnarray}
&&  \langle \phi^{v_1}(\bm{x})| p_x |\phi^{m}(\bm{x}) \rangle
= \frac{i m_t}{\hbar}
\langle \phi^{v_1}(\bm{x})| \left[{\cal H}_{\rm tot}, x \right]_{-}|\phi^{m}(\bm{x}) \rangle
\nonumber\\
&& \qquad  { } = \frac{i m_t}{\hbar} \left( E_1 - E_m \right)\, \langle \phi^{v_1}(\bm{x})|\, x \,|\phi^{m}(\bm{x}) \rangle
\label{EOM} ,
\end{eqnarray}
and similarly for $\langle p_y \rangle$.

Using these relations and the gauge $\bm{A}_{\|}(\bm{r}) = (B_y z, -B_x z, 0)$,
we calculate the matrix elements
\begin{eqnarray}
&& V_{12} = - \frac{|e|}{\hbar} \beta_{D;v_1}\, \cos 2\varphi \, \langle z \rangle   \quad\ \
\label{V12n}\\
&& V_{13} = \left\{ \frac{a_{12}}{\hbar}
\left[ \frac{i m_t}{\hbar} E_{\rm VS} \left( c y_{12} - s x_{12}\right) - B \, \langle z\rangle \right]
\right.
\nonumber\\
&& \qquad
\left. { } + \frac{b_{12}}{\hbar}
\left[ \frac{i m_t}{\hbar} E_{\rm VS}
\left( c x_{12} - s y_{12}\right) - B \sin 2\varphi \, \langle z\rangle \right] \right\} \quad\ \
\label{V13n}\\
&& V_{14} = \left\{ -\frac{a_{12}}{\hbar}
\left[ \frac{i m_t}{\hbar} E_{\rm VS} \left( c x_{12} + s y_{12}\right)  \right]
\right.
\nonumber\\
&& \qquad
\left. { } - \frac{b_{12}}{\hbar}
\left[ \frac{i m_t}{\hbar} E_{\rm VS}
\left( c y_{12} + s x_{12}\right) + |e| B \cos 2\varphi \, \langle z\rangle \right] \right\} \quad\quad
\label{V14n}\\
&& V_{15} = \left\{ \frac{\alpha_{R;v_1}}{\hbar}
\left[ \frac{i m_t}{\hbar} \Delta_{\rm orb} \left( c y_{1,m_1} - s x_{1,m_1}\right) \right]
\right.
\nonumber\\
&& \qquad
\left. { } + \frac{\beta_{D;v_1}}{\hbar}
\left[ \frac{i m_t}{\hbar} \Delta_{\rm orb} \left( c x_{1,m_1} - s y_{1,m_1}\right)  \right] \right\} \quad\ \
\label{V15n}\\
&& V_{16} = \left\{ -\frac{\alpha_{R;v_1}}{\hbar}
\left[ \frac{i m_t}{\hbar} \Delta_{\rm orb} \left( c x_{1,m_1} + s y_{1,m_1}\right)  \right]
\right.
\nonumber\\
&& \qquad
\left. { } - \frac{\beta_{D;v_1}}{\hbar}
\left[ \frac{i m_t}{\hbar} \Delta_{\rm orb} \left( c y_{1,m_1} + s x_{1,m_1}\right)  \right] \right\} \quad\ \
\label{V16n}
\end{eqnarray}
The remaining matrix elements, $V_{17},\ldots , V_{1,12}$ can be obtained from
$V_{15}$, $V_{16}$ by suitable replacements of the envelopes:
$V_{17}=V_{15}(m_1 \to m_2)$, $V_{18}=V_{16}(m_1 \to m_2)$,
$V_{19}=V_{15}(m_1 \to \tilde{m}_1, \Delta_{\rm orb} \to \Delta_{\rm orb}+E_{\rm VS} )$,
$V_{1,10}=V_{16}(m_1 \to \tilde{m}_1, \Delta_{\rm orb} \to \Delta_{\rm orb}+E_{\rm VS})$,
$V_{1,11}=V_{19}(m_1 \to \tilde{m}_2)$, $V_{1,12}=V_{1,10}(m_1 \to \tilde{m}_2)$.
For the second series of matrix elements, they are  related to the above one
(for in-plane magnetic field, $\bm{B}_{\|}$).
Thus, $V_{23}=V_{14}$, $V_{24}=-V_{13}$, $V_{25}=V_{16}$, $V_{26}=-V_{15}$, etc.
$\ldots\, , V_{2,12} = -V_{1,11}$.

Using standard  2nd-order  perturbation theory for the energy
difference  $[\delta E_2^{(2)} - \delta E_1^{(2)}]$  and the above relations
one gets:
\begin{eqnarray}
&& \delta E_2^{(2)} - \delta E_1^{(2)} = \frac{2|V_{12}|^2}{E_Z} +
|V_{14}|^2 \left( \frac{1}{E_Z - E_{\rm VS}} + \frac{1}{E_Z + E_{\rm VS}}\right)
\nonumber\\
&& \quad\quad {} + 2|V_{16}|^2 \left( \frac{1}{E_Z - \Delta_{\rm orb}} + \frac{1}{E_Z + \Delta_{\rm orb}}\right)
\nonumber\\
&& \ \ {} +
2|V_{1,10}|^2 \left( \frac{1}{E_Z - \Delta_{\rm orb}-E_{\rm VS}} + \frac{1}{E_Z + \Delta_{\rm orb}+E_{\rm VS}}\right),
\label{delta-g-parallel-energy}
\end{eqnarray}
%
and for the g-factor one obtains,
by grouping the terms:
\begin{eqnarray}
&&  \delta E_2^{(2)} - \delta E_1^{(2)} = \delta^{(2)}\!g_{\|}^{v_1}\,\, \mu_B B
\nonumber\\
&& \delta^{(2)}\!g_{\|}^{v_1} \equiv  \delta g^{12}_{\|} + \delta g^{14}_{\|} + \delta g^{16}_{\|} 
+ \delta g^{1,10}_{\|} . \quad\ \ \ 
\label{delta-g-parallel-sum}
\end{eqnarray}
The relevant contributions read:
\begin{eqnarray}
&& \delta g^{12}_{\|} = \frac{|e|^2}{\hbar^2 \mu_B^2} \frac{2}{g^*} \beta^2_{D;v_1}\, \cos^2 2\varphi\, \langle z \rangle^2
\label{dg12-parallel}\\
&& \delta g^{14}_{\|} = -\frac{|e|^2}{\hbar^2 \mu_B^2} \frac{1 }{E_{\rm VS}^2-E_Z^2}
\left\{ \frac{2}{g^*}\, E_Z^2\, b_{12}^2\, \cos^2 2\varphi\, \langle z \rangle^2
+ \frac{g^*}{2} E_{\rm VS}^2
\right.
\nonumber\\
&& \left. { }   
\qquad  \times  \frac{m_t^2}{m_0^2}   
\left[ a_{12} \left( c x_{12} + s y_{12}\right)  + b_{12} \left( c y_{12} + s x_{12}\right)\right]^2 \right\}
\label{dg14-parallel}\\
&& \delta g^{16}_{\|}  + \delta g^{18}_{\|} =
-\frac{|e|^2}{\hbar^2 \mu_B^2} \frac{g^*}{2} \frac{m_t^2}{m_0^2} \frac{\Delta^2_{\rm orb}}{\Delta^2_{\rm orb} - E_Z^2}
\nonumber\\
&&
\times
\left\{  y_{1,m_1}^2 \left[ s \, \alpha_{R;v_1} + c \, \beta_{D;v_1} \right]^2
 + x_{1,m_2}^2 \left[ c \, \alpha_{R;v_1} + s \, \beta_{D;v_1} \right]^2   \right\} \quad\
\label{dg1618-parallel}\\
&& \delta g^{1,10}_{\|}  + \delta g^{1,12}_{\|} =
-\frac{|e|^2}{\hbar^2 \mu_B^2} \frac{g^*}{2}  \frac{m_t^2}{m_0^2}     
\frac{\left(\Delta_{\rm orb} + E_{\rm VS} \right)^2}{\left(\Delta_{\rm orb} + E_{\rm VS} \right)^2-E_Z^2}
\nonumber\\
&& \quad
\times
\left\{ y_{1,\tilde{m}_1}^2 \left[ s \, a_{12} + c \, b_{12} \right]^2
+ x_{1,\tilde{m}_2}^2 \left[ c \, a_{12} + s \, b_{12} \right]^2   \right\} .
\label{dg1012-parallel}
\end{eqnarray}
In the above,
we have used  (for a circular dot with parabolic confinement)  that:
$x_{1,m_1} = y_{1,m_2} = x_{1,\tilde{m}_1} = y_{1,\tilde{m}_2} = 0$.
The standard non-zero  dipole matrix elements to orbital states,
$y_{1,m_1}=x_{1,m_2} = y_{1,\tilde{m}_1} = x_{1,\tilde{m}_2} = \sqrt{\frac{\hbar^2}{2 m_t \Delta_{\rm orb}}}$
will be used for further evaluation of
Eqs.~(\ref{dg1618-parallel})  and (\ref{dg1012-parallel}).

\subsection{Second order corrections: case of  $\bm{B}_{\perp}$}
\label{App B: B-perpendicular}

For the second order corrections in perpendicular
magnetic field $\bm{B}_{\perp}$ we use the SOC Hamiltonians
Eq.~(\ref{spin-valley-B-field})
and include transitions to higher states as was done above.
One obtains for the
first few matrix elements
\begin{eqnarray}
&& V_{12} = \langle \bar{v}_1, \downarrow |{\cal H}_{\rm s-v}|\bar{v}_1, \uparrow \rangle
\nonumber\\
&& { } = \hbar^{-1}\langle \phi^{v_1}(\bm{x})| a_{11} \left( P_y - i P_x \right) +
b_{11} \left( P_x - i P_y \right)|\phi^{v_1}(\bm{x}) \rangle \qquad
\label{V12-perp}\\
&& V_{13} = 0
\label{V13-perp}\\
&& V_{14} = \langle \bar{v}_1, \downarrow |{\cal H}_{\rm s-v}|\bar{v}_2, \uparrow \rangle
\nonumber\\
&& { } = \hbar^{-1}\langle \phi^{v_1}(\bm{x})| a_{12} \left( P_y - i P_x \right) +
b_{12} \left( P_x - i P_y \right)|\phi^{v_2}(\bm{x}) \rangle \qquad
\label{V14-perp} \\
&& V_{15} = 0
\label{V15-perp}\\
&& V_{16} = \langle \bar{v}_1, \downarrow |{\cal H}_{\rm s-v}|m_1, \uparrow \rangle
\nonumber\\
&& { } = \hbar^{-1}\langle \phi^{v_1}(\bm{x})| a_{11} \left( P_y - i P_x \right) +
b_{11} \left( P_x - i P_y \right)|\phi^{m_1}(\bm{x}) \rangle , \qquad
\label{V16-perp}
\end{eqnarray}
etc.
The structure of the higher matrix elements is similar, e.g.,
$V_{17} = V_{19} = V_{1,11} = 0$,
$V_{18} = V_{16}(m_1 \to m_2)$, $V_{1,10} = V_{16}(m_1 \to \tilde{m}_1)$,
 $V_{1,12} = V_{16}(m_1 \to \tilde{m}_2)$.
For the second series of matrix elements, they are related to the above one
(for perpendicular magnetic field, $\bm{B}_{\perp}$)
Thus,
$V_{23}= V_{14}(i \to -i)$,
$V_{25}= V_{16}(i \to -i)$, $V_{27}= V_{18}(i \to -i)$,
$V_{29}= V_{1,10}(i \to -i)$,
$V_{2,11}= V_{1,12}(i \to -i)$.
For the squared matrix elements, these replacements correspond
to the formal sign change of $E_Z = g^* \mu_B B$ (see below).

Using standard  2nd-order  perturbation theory for the energy
difference  $[\delta E_2^{(2)} - \delta E_1^{(2)}]$  and the above relations
one gets:
%
\begin{eqnarray}
&& \delta E_2^{(2)} - \delta E_1^{(2)} = \frac{2|V_{12}|^2}{E_Z} +
 \left( \frac{|V_{23}|^2}{E_Z - E_{\rm VS}} + \frac{|V_{14}|^2}{E_Z + E_{\rm VS}}\right)
\nonumber\\
&&
{} + \left( \frac{|V_{25}|^2}{E_Z - \Delta_{\rm orb}} + \frac{|V_{16}|^2}{E_Z + \Delta_{\rm orb}}\right)
+ (m_1 \to m_2)
\nonumber\\
&& 
{} + \left( \frac{|V_{29}|^2}{E_Z - \Delta_{\rm orb}-E_{\rm VS}} + \frac{|V_{1,10}|^2}{E_Z + \Delta_{\rm orb}+E_{\rm VS}}\right)
\nonumber\\
&& { } + (\tilde{m}_1 \to \tilde{m}_2)
\label{delta-g-perp-energy}  .
\end{eqnarray}

The matrix elements $V_{ab},\ a=1,2,\,b=1,\ldots,12$, are calculated
similar to the previous case,  using the equation of motion, Eq.~(\ref{EOM}).

Having at hand these matrix elements,  we use the
2nd-order correction to the energy difference, Eq.~(\ref{delta-g-perp-energy}),
and group the terms accordingly:
\begin{eqnarray}
&&  \delta E_2^{(2)} - \delta E_1^{(2)} = \delta^{(2)}\!g_{\perp}^{v_1}\,\, \mu_B B
\nonumber\\
&& \delta^{(2)}\!g_{\perp}^{v_1} \equiv  \delta g^{12}_{\perp} + \delta g^{14}_{\perp} + \delta g^{16}_{\perp} + \delta g^{18}_{\perp}
+ \delta g^{1,10}_{\perp} + \delta g^{1,12}_{\perp} . \quad\ \ \
\label{delta-g-perp-sum}
\end{eqnarray}

The relevant contributions to $\delta^{(2)}\!g_{\perp}$ read:
\begin{eqnarray}
&& \delta g^{12}_{\perp} = \frac{|e|^2}{\hbar^2 \mu_B^2} \frac{1}{2 g^*}
\left\{ \left( x_{11} \alpha_{R;v_1} - y_{11} \beta_{D;v_1} \right)^2  \right.
\nonumber\\
&& \qquad\qquad\qquad\quad\quad
\left. { } +
\left( x_{11} \beta_{D;v_1} - y_{11} \alpha_{R;v_1} \right)^2
  \right\} \qquad\quad
\label{dg12-perp}
\end{eqnarray}
which coincides with Eq.~(\ref{dg2-perp-1st}), as expected. Also,
\begin{eqnarray}
&& \delta g^{14}_{\perp} =
\frac{g^*}{E_Z} \left[\frac{|V_{14}|^2}{E_Z + E_{\rm VS}} + \frac{|V_{23}|^2}{E_Z - E_{\rm VS}} \right]
 = \frac{|e|^2}{4 \hbar^2 \mu_B^2} \frac{1 }{h_z \left( \frac{m_0}{m_t} + 2 h_z \right)}
\nonumber\\
&& \qquad
\times
\left\{
\left[ x_{12}\, a_{12} (1- h_z) + y_{12}\, b_{12} (1+ h_z)\right]^2 \right.
\nonumber\\
&&\qquad\qquad\
\left. { } + \left[ x_{12}\, b_{12} (1+ h_z) + y_{12}\, a_{12} (1- h_z)\right]^2 \right\}
\nonumber\\
&&\qquad  { } + (h_z \to -h_z)
\label{dg14-perp}
\end{eqnarray}
with  $h_z \equiv \frac{m_0}{m_t}\frac{E_Z}{2 E_{VS}}$.
\begin{eqnarray}
&& \delta g^{16}_{\perp} + \delta g^{18}_{\perp}=
\frac{g^*}{E_Z} \left[\frac{|V_{16}|^2}{E_Z + \Delta_{\rm orb}} + \frac{|V_{25}|^2}{E_Z - \Delta_{\rm orb}} \right]
+ (m_1 \to m_2) \qquad
\nonumber\\
&& \qquad\qquad\ \ \
{ } = \frac{|e|^2}{4 \hbar^2 \mu_B^2} \frac{1 }{\tilde{h}_z \left( \frac{m_0}{m_t} + 2 \tilde{h}_z \right)}
\left\{
 \left( y_{1,m_1}^2 + x_{1,m_2}^2 \right) \right.
\nonumber\\
&&  \qquad\qquad\qquad
\left.
\times \left( \alpha_{R;v_1}^2 (1- \tilde{h}_z)^2 + \beta_{D;v_1}^2 (1+ \tilde{h}_z)^2\right) \right\}
\nonumber\\
\nonumber\\
&&\qquad\qquad\quad  { } + (\tilde{h}_z \to -\tilde{h}_z)
\label{dg1618-perp}
\end{eqnarray}
with  $\tilde{h}_z \equiv \frac{m_0}{m_t}\frac{E_Z}{2 \Delta_{\rm orb}}$.
\begin{eqnarray}
&& \delta g^{1,10}_{\perp} + \delta g^{1,12}_{\perp}=
\frac{g^*}{E_Z} \left[\frac{|V_{1,10}|^2}{E_Z + \Delta_{\rm orb} + E_{\rm VS}} + \frac{|V_{29}|^2}{E_Z - \Delta_{\rm orb} - E_{\rm VS}} \right]
\nonumber\\
&& \
{ } + (\tilde{m}_1 \to \tilde{m}_2) = \frac{|e|^2}{4 \hbar^2 \mu_B^2} \frac{1 }{\bar{h}_z \left( \frac{m_0}{m_t} + 2 \bar{h}_z \right)}
\left\{
 \left( y_{1,\tilde{m}_1}^2 + x_{1,\tilde{m}_2}^2 \right) \right.
\nonumber\\
&&  \qquad\qquad\qquad\quad
\left.
\times \left( a_{12}^2 (1- \bar{h}_z)^2 + b_{12}^2 (1+ \bar{h}_z)^2\right) \right\}
\nonumber\\
&&\qquad\qquad\qquad  { } + (\bar{h}_z \to -\bar{h}_z)
\label{dg1012-perp}
\end{eqnarray}
with  $\bar{h}_z \equiv \frac{m_0}{m_t}\frac{E_Z}{2 \left(\Delta_{\rm orb} + E_{\rm VS} \right)}$.
In the above,
we have used the relations for the dipole matrix elements to orbital states,
see text after
Eqs.~(\ref{dg1618-parallel})  and (\ref{dg1012-parallel}).
%

As mentioned above, for an interface with roughness the lowest energy envelopes,
$\phi^{v_i}(x,y)$
(quasi s-like) acquire a p-like contribution, depending on the eigenvalley index $v_i$.
Thus, the dipole matrix elements
$\bm{r}_{ij} \equiv \langle v_i| \bm{r} |v_j\rangle$ $i,j = 1,2$
are generally non-zero \cite{Yang2013NC,Gamble2013PRB},
getting size of few nm for this type of QDs\cite{Yang2013NC}.

\subsection{$\delta g$ at the spin-valley anticrossing point}
\label{App B:  delta g at anticrossing}

At the anticrossing (at the so-called ``relaxation hot spot'')\cite{Yang2013NC},
when $E_Z \approx E_{\rm VS}$,  the contribution
$\delta g^{14}_{\|}$ acquires a first order correction
(by solving the standard secular equation).
The exact qubit energy difference is
$\frac{1}{2}\left[ E_{\rm VS}+ E_Z - \sqrt{\left( E_{\rm VS} - E_Z \right)^2 + \Delta_a^2} \right]$,
where
\begin{equation}
\Delta_a = 2 |V_{23}| = 2 |V_{14}|
\label{delta-anticros}
\end{equation}
is the splitting at anticrossing of the relevant valley states\cite{Yang2013NC,Xiao2014NC}
$|2\rangle$ and $|3\rangle$, see Eqs.~(\ref{V14n}), (\ref{V14-perp}), and Fig.~\ref{fig:1}d.
%
Close to anticrossing, when   $\delta \equiv E_{\rm VS} - E_Z \lesssim \Delta_a$,
\begin{equation}
\delta g_{\rm hot-spot} = - \frac{\Delta_a}{E_Z} + \frac{\delta}{E_Z} - \frac{\delta^2}{2\Delta_a E_Z} .
\label{delta-g-anticros}
\end{equation}
Thus $\delta g_{\rm hot-spot}$ may be  of the order of $10^{-3}$ or less since the splitting
was evaluated \cite{Yang2013NC,Xiao2014NC} as
$\Delta_a = (10^{-3} - 10^{-4})\, E_{\rm VS}$.
This is at least $~10$ times smaller than the observed experimental $g$-factor
correction \cite{Veldhorst2014Nn,VeldhorstRuskov2015PRB},
as presented on Fig.~\ref{fig:3}.
Also, there is no any observed deviation from the linear dependence with $F_z$
near the anticrossing point which restricts the size of $\Delta_a$.

\subsection{The integral relation, Eq.~(\ref{integral_relation}),
for a $z$-confinement potential $U(z)$ with an infinite boundary}
\label{App B: integral relation}

One starts with the one-dimensional eigenvalue problem
\begin{equation}
-\frac{\hbar^2}{2 m} \varphi''(z) + U(z) \varphi(z) - E \varphi(z) = 0
\label{1D-problem}
\end{equation}
with $\varphi(0) = 0$.   
By multiplying Eq.~(\ref{1D-problem}) by ${\varphi^{*}}'(z)$ and integrating by parts
the first and last term:
\begin{eqnarray}
&& \int_0^{\infty} dz {\varphi^{*}}^{'}(z) \varphi''(z)
= -{\varphi^{*}}'(0) \varphi'(0) - \int_0^{\infty} dz {\varphi^{*}}''(z) \varphi'(z) \qquad\ \
\\
&& - E\, \int_0^{\infty} dz {\varphi^{*}}'(z) \varphi(z) = E\, \int_0^{\infty} dz \varphi^*(z) \varphi'(z) ,
\label{by-parts}
\end{eqnarray}
then one adds the conjugate 1D-equation, multiplied by $\varphi'(z)$.
As a result, $- \int_0^{\infty} dz \frac{2 m}{\hbar^2} U(z) \frac{d}{dz} |\varphi(z)|^2 = |\varphi'(0)|^2$
or
\begin{equation}
\frac{2 m}{\hbar^2} \, \int_0^{\infty} dz  \varphi^*(z) \partial_z U(z) \varphi(z) = |\varphi'(0)|^2
\label{integral_relaton_app} .
\end{equation}

\section{Interface boundary condition from hermiticity of the Hamiltonian}
\label{App C: BC and hermiticity}

\subsection{Volkov-Pinsker boundary condition}
\label{App C: Volkov-Pinsker}

For completeness, we first re-derive the Volkov-Pinsker BC \cite{VolkovPinsker1979}, starting from
single-band approximation Hamiltonian, in the presence of external field, $\bm{A}(\bm{r})$:
\begin{equation}
{\cal H} = \frac{(\bm{p} + |e| \bm{A})^2}{2 m} + U(\bm{r})
\label{single-band-A} .
\end{equation}
Considering two arbitrary solutions, $\phi_1$, $\phi_2$ of the Schr\"{o}dinger equation,
one states the hermiticity condition at the half-space, $z > z_0$ \cite{VolkovPinsker1979}:
\begin{equation}
\int_{z > z_0} dz \, \phi_1^{\dagger} \left({\cal H} \phi_2 \right)  =
\int_{z > z_0} dz \, \phi_2 \left({\cal H} \phi_1 \right)^*
\label{hermiticity} .
\end{equation}
Substituting ${\cal H}$ in Eq.~(\ref{hermiticity},) and integrating by parts
one gets the relation (put $\hbar = e = 1$):
\begin{equation}
\phi_2(z_0) \frac{d\phi_1^{*}}{dz}  - \phi_1^{*}(z_0) \frac{d\phi_2}{dz}
+ 2 i \phi_1^{*}(z_0) A_z(z_0) \phi_2(z_0) = 0
\label{hermiticity-BC1} ,
\end{equation}
where  separation of variables is assumed for the potential, Eq.(\ref{conf_potentials-z}).
Eq.~(\ref{hermiticity-BC1}) can be satisfied if
\begin{equation}
\frac{1}{\phi_1} \frac{d\phi_1(z_0)}{dz} =
\frac{1}{\phi_2} \frac{d\phi_2(z_0)}{dz} = {\it const.} + i A_z(z_0)
\label{hermiticity-BC1-solution-of} .
\end{equation}
By choosing  ${\it const.} \equiv -\frac{1}{R}$  one can recast
Eq.~(\ref{hermiticity-BC1-solution-of}) to the BC:
\begin{equation}
\left\{ 1 + i \frac{R}{\hbar} \left( p_z + |e| A_z \right) \right\}\,\phi(z) \mid_{z_0^{+}} = 0
\label{VolkovPinskerBC-A} ,
\end{equation}
with $p_z \equiv -i \hbar \partial_z$.
For $A_z = 0$ one recovers Eq.(\ref{VolkovPinskerBC1}).
%
As follows from Eq.~(\ref{VolkovPinskerBC-A}), the gauge invariance of the Schrodinger
equation plus boundary conditions implies in general ``extension of derivatives'' both in the Hamiltonian
and in the boundary conditions. In case of the spin-valley BCs considered in the main text,
Eqs.~(\ref{BC}), (\ref{interface_H}),  and (\ref{BC-spinor}),
one should extent both the $\partial_z$-derivative as well as the $\partial_{x,y}$-derivatives.

Notice also that the bulk velocity operator is
$v_z \equiv \frac{\partial {\cal H}}{\partial p_z} = \frac{1}{m}\left( p_z + |e| A_z \right)$.
The hermiticity condition, Eq.~(\ref{hermiticity-BC1}) then can be rewritten as
\begin{equation}
\phi_1^{*} \left(v_z \phi_2 \right) + \left(v_z \phi_1 \right)^* \phi_2\,  \mid_{z_0^{+}} = 0
\label{hermiticity-BC2-velocity} .
\end{equation}
This  implies  continuity of the envelope flux density, despite of the discontinuity of the
wave function at its derivative at the interface.

\subsection{BC and gauge-invariance}
\label{App C: Gauge-invariance}

Concerning  the gauge invariance, we have already mentioned in Sec. \ref{sec: Sec IV: g-factor correction}
that the problem (Hamiltonian plus boundary conditions) is written in a gauge invariant form,
via the extension of the derivatives.
Therefore, in the actual calculations, one is using the most convenient gauge
as is, e.g., with the results for the $g$-factor renormalization,
Eqs.~(\ref{H_B_parallel})-(\ref{g-factor-change}).
One may ask the question how the gauge invariance is preserved during the derivation, e.g., of Eq.~(\ref{g-factor-change})?
One mention that any gauge change leads to a multiplication of the wave function
with a phase factor, which cancels in the quantum average $\langle z\rangle$
in Eq.~(\ref{g-factor-change})
[considering a boundary at $z_0=0$].
By using  the gauge $\bm{A}_{\|}(\bm{r}) = (z, -z, 0) B/\sqrt{2}$,
for each of the two spin components,
there is a modification of the $z$-confinement potential of Eq.~(\ref{conf_potentials-z}) by a linear $z$-term.
This leads to a modification of the eigenvalues of the original problem, Eq.~(\ref{bulk_H}), which ends up
with the result Eq.~(\ref{g-factor-change}) as a first order correction.
Since we are considering a homogeneous magnetic field, the vector potential is a linear function of the coordinates,
including also an arbitrary constant vector.
E.g., for the gauge $\bm{A}' = \bm{A}_{\|}(\bm{r}) + (c,-c,0)$ one naively would expect a shift in the $z$-coordinate.
This  gauge transformation, however, corresponds to adding a constant to the
Hamiltonian Eq.~(\ref{bulk_H}), which  does not change the eigenvalues.
Thus, the gauge invariance is preserved in this case.

One may consider the gauge $\bm{A}'' = (0,0,y-x) B / \sqrt{2}$,  which is more involved.
Indeed, in this case there is no explicit $z$, and one is puzzling how one can obtain the $\langle z\rangle$ in the final result.
One starts with the BC, Eq.~(\ref{BC}) in the form
\begin{eqnarray}
&& \left\{ 1 +
i R (k_z + \frac{|e|}{\hbar} A_z) - R\, \frac{2m_l}{\hbar^2}\, V_{\rm if}(\bm{k}) \right\} \Phi(\bm{r})\mid_{z=0^{+}}
\nonumber\\
&& \qquad\qquad\qquad\qquad\qquad \equiv {\cal B}(A_z) \Phi(\bm{r})\mid_{z=0^{+}} = 0
\label{BCy-x}  ,
\end{eqnarray}
and following the derivations of
Eqs.~(\ref{transformed-H})  and (\ref{eff-interface-Hamiltonian}),
one obtains the effective unitary transform (see Appendix \ref{App A: eff-Hamiltonian})
\begin{equation}
\Gamma_{BC}(A_z) =
1 + i \left[R (k_z + \frac{|e|}{\hbar} A_z) + R^2 \frac{2m_l}{\hbar^2} V_{\rm if}(\bm{k}) (k_z + \frac{|e|}{\hbar} A_z)\right]
\label{unitary transform-y-x}
\end{equation}
%
such that $\Gamma_{BC}(A_z) \Phi(\bm{r})\mid_{z=0^{+}} \simeq 0 $.
After some elaborate calculation, using the above described procedure, one can obtain a term in
the effective Hamiltonian perturbation, $\Delta H$, which is $~ k_z^2$.
Thus,
since $\langle k_z^2 \rangle = {\it const.} \langle z \rangle$
for the triangular potential in Eq.~(\ref{bulk_H}), and $\langle z \rangle$ is recovered.

\subsection{Estimation of the R parameter}
\label{App C: R-parameter}

One can illustrate how an effective length parameter $R$ appears in a
single-band
BC like Eq.~(\ref{VolkovPinskerBC-A})
from a two-band model\cite{VolkovPinsker1979},
with 2-component envelope, $\phi^T = [\phi_c(\bm{r}),\phi_v(\bm{r})]$,
including conduction and valence bands.
Neglecting ${\cal O}(p_z^2)$ effects,
the $\bm{k}\cdot\bm{p}$-Hamiltonian is
\begin{equation}
{\cal H}^{\rm 2band} =
\left(
\begin{array}{cc}
E_c & \frac{p_{\rm cv}}{m_0}\, p_z \\
\frac{p^*_{\rm cv}}{m_0}\, p_z & E_v
\end{array}
\right) ,
\label{2-band}
\end{equation}
where $p_{\rm cv}$ is the interband momentum matrix element.
The BC, Eq.~(\ref{hermiticity-BC2-velocity}), is recasted to
$(-\phi_{1v}^*\, \phi_{2c} + \phi_{1c}^*\, \phi_{2v} )\mid_{z_0^{+}} = 0$,
for any two functions, $\phi_{1}$, $\phi_{2}$.
On the other hand, a stationary solution of the Schr\"{o}dinger equation
with ${\cal H}^{\rm 2band}$
gives a relation: $\phi_c = - \frac{p_{\rm cv} p_z}{(E_c - E)m_0} \phi_v$
(and analogous one, with $c \to v$), allowing to exclude the other band.
(It is worth to stress here, that such relations make it impossible to have
simultaneously $\phi_c(z_0)=0$ and $\phi_v(z_0)=0$, as required by the standard BC
with infinite boundary).
Compatibility of the two-band BC with the single-band BC, Eq.~(\ref{VolkovPinskerBC-A}),
leads to the relation\cite{VolkovPinsker1979}:
$R_c \simeq R_v =
 \frac{1}{2}\,\sqrt{\frac{2 \hbar^2}{m_c^* E_{\rm gap} }}$,
where $E_{\rm gap} = E_c - E_v \approx 4\, {\rm eV}$ is the band gap in Si
at the band minima,
$m_c^*$
is the effective mass,
and we have used the approximate relation \cite{YuCardonaBook}:
$\frac{m_0}{m_c^*} \approx \frac{2 p_{\rm cv}^2}{m_0 E_{\rm gap}}$.
Thus, as a rough estimation (i.e., not taking into account valleys)
one gets $R = R_c^{\rm Si} \approx 0.1 - 0.2\, {\rm nm}$
for $m_t < m_c^* < m_l$.


%

\end{document}